\documentclass[fleqn,usenatbib]{mnras}

\usepackage{newtxtext,newtxmath}

\usepackage[T1]{fontenc}

\DeclareRobustCommand{\VAN}[3]{#2}
\let\VANthebibliography\thebibliography
\def\thebibliography{\DeclareRobustCommand{\VAN}[3]{##3}\VANthebibliography}

\usepackage{graphicx}	
\usepackage{amsmath}	
\usepackage{pifont}
\usepackage{mathtools}
\usepackage{placeins}
\usepackage{float}
\usepackage{bm}
    
\usepackage{nccmath}

\usepackage{cleveref}
\crefname{figure}{Fig.}{Figs.}
\crefname{table}{Table}{Tables}

\title[Non-ideal magnetohydrodynamics on a moving mesh]{Non-ideal magnetohydrodynamics on a moving mesh I: Ohmic and ambipolar diffusion}

\author[O. Zier, V. Springel and A. C. Mayer]{%
Oliver Zier$^{1}$\thanks{E-mail: ozier@mpa-garching.mpg.de}, 
 Volker Springel$^{1}$ and Alexander C. Mayer$^{1}$
\vspace*{0.1cm}\\%
$^{1}$Max-Planck-Institut für Astrophysik, Karl-Schwarzschild-Straße 1, 85741 Garching, Germany\\
}

\date{Accepted XXX. Received YYY; in original form ZZZ}

\pubyear{2023}

\begin{document}
 \label{firstpage}
\pagerange{\pageref{firstpage}--\pageref{lastpage}}
\maketitle

\begin{abstract}
Especially in cold and high-density regions, the assumptions of ideal magnetohydrodynamics (MHD) can break down, making first order non-ideal terms such as Ohmic and ambipolar diffusion as well as the Hall effect important. In this study we present a new numerical scheme for the first two resistive terms, which we implement in the moving-mesh code {\small AREPO} using the single-fluid approximation combined with a new gradient estimation technique based on a least-squares fit per interface. Through various test calculations including the diffusion of a magnetic peak, the structure of a magnetic C-shock, and the damping of an Alfvén wave, we show that we can achieve an accuracy comparable to the state-of-the-art code {\small ATHENA++}. We apply the scheme to the linear growth of the magnetorotational instability and find good agreement with the analytical growth rates. By simulating the collapse of a magnetised cloud with constant magnetic diffusion, we show that the new scheme is stable even for large density contrasts. Thanks to the Lagrangian nature of the moving mesh method the new scheme is thus well suited for intended future applications where a high resolution in the dense cores of collapsing protostellar clouds needs to be achieved. In a forthcoming work we will extend the scheme to  the Hall effect.
\end{abstract}

\begin{keywords}
methods: numerical -- MHD -- instabilities -- dynamo -- turbulence
\end{keywords}

\section{Introduction}

Magnetic fields play an important role on various scales in the Universe. They can be amplified by dynamo effects and can decisively influence the evolution of conducting fluids \citep{ferriere2001interstellar, cox2005three}, as well as the evolution of relativistic, charged particles, typically called cosmic rays \citep{fermi1949origin, kotera2011astrophysics}. 

Computer simulations are a powerful tool for understanding this interplay, and more generally, for describing the evolution of complex non-linear systems in astrophysics. To compute the joint dynamics of gas flows and magnetic fields, one can extend the Euler equations, describing an ideal fluid, by assuming a perfect coupling of the fluid with the magnetic field. Neglecting resistive effects leads to the ideal MHD approximation, which has been used in simulations of large cosmological boxes \citep[e.g.][]{marinacci2015large,marinacci2015effects, dolag2016sz, marinacci2018first}, galaxy clusters \citep[e.g.][]{dolag1999sph, dolag2002evolution}, the magnetic field within individual galaxies \citep[e.g.][]{pakmor2014magnetic, pakmor2017magnetic}, the interstellar medium \citep[e.g.][]{kim2018numerical, simpson2023cosmic}, molecular clouds \citep[e.g.][]{grudic2021starforge} and the collapse of protostellar cores \citep[e.g.][]{mellon2008magnetic}.

Although the ideal MHD approximation can be justified by a high ionization rate of the fluid in most of these examples, at small scales the ideal MHD approximation can break down. To account for this, one must add non-ideal MHD effects such as Ohmic and ambipolar diffusion, as well as the Hall effect.

Ambipolar diffusion enables a decoupling of neutral gas from the magnetic field, which is crucial for star formation \citep{basu2004formation}. It can also resolve the so-called fragmentation crisis by reducing the magnetic pressure stabilizing overdensities in molecular cloud cores \citep{hennebelle2008magnetic}. Ambipolar diffusion can influence MHD turbulence by steepening the velocity and magnetic field power spectrum \citep{li2008sub} and by altering the velocity and density structure of the gas \citep{ntormousi2016effect}.

Ohmic diffusion is essential for magnetic reconnection, which describes  changes in the topology of the magnetic field. While reconnection is suppressed in ideal MHD by the flux freezing condition, this phenomenon can lead to strong heating at the point of reconnection through Joule heating, which is thought to be important, e.g, in the Solar corona where it fuels eruptive events \citep{parker1983magnetic, Cheng2017}. Additionally, Ohmic diffusion can reduce the decay time of MHD turbulence in molecular clouds \citep{basu2010long}.

The collapse of protostellar cores and the evolution of protostellar disks are prime examples for the influence of non-ideal MHD effects. While purely hydrodynamic simulations show the formation of a rotationally supported disk during cloud collapse due to the conservation of angular momentum \citep{Bate2011,bate2014collapse}, ideal MHD can lead to the so-called magnetic braking catastrophe, which prevents disk formation by significantly increasing the transport of angular momentum \citep[see e.g.][]{Allen2003}.
Non-ideal effects can reduce the magnetic field strength and, therefore, the efficiency of angular momentum transport. This in turn can lead to the formation of a rotationally supported disk \citep{wurster2016can}. For a recent review of the influence of magnetic fields on the formation of protostellar disks, we refer to \cite{wurster2018role}.

After the formation of a protostellar disk, an effective viscosity is necessary to explain the transport of angular momentum \citep{shakura1973black,lynden1974evolution}. Viscous behaviour can for example be generated by turbulence resulting from instabilities. In particular, the magnetorotational instability (MRI) is a promising candidate for explaining the origin of the effective viscosity in ideal MHD, as it only requires a small seed field to grow in Keplerian-like shear flows \citep{velikhov1959stability, chandrasekhar1960stability, fricke1969stability, balbus1991powerful}. The MRI is often studied in the shearing box approximation, which enables a high-resolution analysis of a small portion of the disk, reducing the computational costs \citep{hill1878collected,goldreich1965ii}. This is crucial for large parameter studies, which are necessary to explore the extensive parameter space of non-ideal MHD. However, non-ideal MHD effects can slow down the linear growth of the MRI, and even suppress it entirely \citep{lesur2021magnetohydrodynamics}. While Ohmic diffusion only suppresses the MRI, ambipolar diffusion can increase the growth rate of oblique modes, which have a non-vanishing wave number in radial direction \citep{kunz2004ambipolar}. Non-ideal MHD effects can also impact the formation of disk winds in this scenario \citep[e.g.][]{bai2014hall}, making it essential to understand their influence to comprehend the evolution of protostellar and protoplanetary disks.

Due to the significance of non-ideal MHD effects, it is unsurprising that they have been incorporated into various numerical codes. The most general technique is the multi-fluid approach, which involves separately solving the momentum and continuity equations for critical species, including, e.g., ions, electrons, and neutrals \citep[e.g.,][]{inoue2008two,inoue2009two}.  The species can interact with each other via coupling terms. If one assumes that the mass density is dominated by the mass density of neutrals, and that collisions are dominated by interactions between neutrals and charged particles, one can neglect the inertia and pressure of charged particles as well as the collision between them \citep{o2006explicit, rodgers2016global}. This approximation requires the solution of individual continuity equations for each species but only one momentum equation for the neutrals. 

By using the strong coupling approximation, which means the magnetic field as well as neutral flows evolve on longer time scales than charged particles and one can neglect the pressure and momentum of ions in comparison to neutrals, one can combine the individual continuity equations into a single one for the total mass density \citep[see e.g.][]{choi2009explicit}. We will concentrate in this paper on this single-fluid approximation which was implemented in Eulerian codes such as {\small PLUTO} \citep{lesur2014thanatology}, {\small ATHENA} \citep{bai2014hall}, {\small ATHENA++} \citep{stone2020athena++}, {\small RAMSES} \citep{masson2012incorporating, marchand2018impact}, {\small ZeusTW} \citep{li2011non} and {\small MANCHA3D} \citep{gonzalez2018mhdsts}, as well as Lagrangian particle methods such as the SPH code by \cite{tsukamoto2017impact}, {\small PHANTOM} \citep{price2018phantom} and {\small GIZMO} \citep{hopkins2017anisotropic}. Although both types of methods are widely used, Eulerian codes typically suffer from large advection errors in regions with large bulk velocities, whereas the smoothed particle hydrodynamics (SPH) method yields noisier results of lower accuracy but offers an automatically adaptive resolution.

The moving mesh method \citep{springel2010pur,weinberger2020arepo} is a comparatively new Lagrangian approach that tries to combine the advantages of a Galilei-invariant Lagrangian method with the high accuracy of the finite volume methods typically employed in Eulerian codes. This makes it especially interesting for global disk simulations but also for local simulations with large density gradients that can benefit from the code's high flexibility to continuously adapt cell sizes, and to increase and decrease the local resolution by splitting and merging individual computational cells. In \cite{ZierMRI} we showed that the code can resolve the magnetorotational instability (MRI) in the case of ideal MHD with similar accuracy as static grid codes.

\cite{marinacci2018non} already implemented both an explicit and an implicit scheme for Ohmic diffusion in {\small AREPO}. However, to our knowledge this implementation has not been utilized yet in science applications beyond demonstrating second-order convergence. The implementation only computed the normal component of the magnetic field gradients on the interfaces between adjacent cells using a finite difference method, which was sufficient for calculating the fluxes for Ohmic diffusion. However, for ambipolar diffusion and the Hall effect, the missing components of the gradient are also required, making it difficult to generalize the scheme to these additional non-ideal effects. In the present work, we have thus modified the method presented in \cite{pakmor2016semi} to estimate the complete gradient of the magnetic field at cell interfaces. By utilizing these gradients, we have developed an explicit solver for both Ohmic and ambipolar diffusion, and we plan to extend it to the Hall effect in the future.

The paper is structured as follows. In Section~\ref{sec:numericalMethods}, we introduce the non-ideal MHD equations with Ohmic and ambipolar diffusion. We further describe the implementation of ideal MHD in {\small AREPO} and our new scheme for solving the additional non-ideal terms. In Section~\ref{sec:Tests}, we test our new module by applying it to several test problems with known analytical or semi-analytical solutions. This includes the diffusion of a magnetic field pulse, the structure of a C-shock modified by ambipolar and Ohmic diffusion, and the damping of Alven waves. In all cases, we check the convergence of our scheme with increasing resolution and compare our results with the state-of-the-art code {\small ATHENA++}, which solves the same physical equations on a static grid. In Section~\ref{sec:linearMRI} we introduce the shearing box approximation as implemented in \cite{zier2022simulating} in the {\small AREPO} code. We describe the linearised equations for the magnetorotational instability with a net vertical magnetic field and ambipolar as well as Ohmic diffusion. We measure the growth rates of the magnetic field during the exponential growth as a function of the strength of the non-ideal effects and compare it with those obtained with {\small ATHENA++} as well as with analytical results. In Section~\ref{sec:magnetizedCloudCollapse}, we qualitatively analyse the impact of non-ideal MHD effects on the gravitational collapse of a magnetized gas cloud to show that our new solver works accurately even in more complicated setups. Finally, in Section~\ref{sec:discussionSummary} we discuss and summarise our results.

\section{Numerical methods}
\label{sec:numericalMethods}

\subsection{The non-ideal MHD equations}

The non-ideal MHD equations in conservative form can be written as:
\begin{equation}
    \frac{\partial \bm U }{\partial t}+ \nabla \cdot \bm F_{\rm ideal}(\bm U) +  \nabla \cdot \bm F_{\rm nonid}(\bm U)=0.
    \label{eq:nonidealMHDEquations}
\end{equation}
Here, we introduced the state vector $\bm U$, the flux function $\bm F_{\rm ideal}$ describing the effect of ideal MHD, and the flux function $\bm F_{\rm nonid}$ for the non-ideal MHD effects of Ohmic and ambipolar diffusion. They are given in Heaviside–Lorentz units (which we exclusively use in this paper with the exception of Section~\ref{sec:magnetizedCloudCollapse}) by:
\begin{align}
\bm U  = \begin{pmatrix}
   \rho \\
   \rho \bm v  \\
   \rho e  \\
   \bm B\\
   \end{pmatrix},  \;\;\;\;\;\; 
   \bm F_{\rm ideal}(\bm U) =  \begin{pmatrix}
   \rho \bm v\\
   \rho \bm v \bm v^T + P -\bm B \bm B^T\\
   \rho e  \bm v + P \bm v -\bm B(\bm v \cdot \bm B)\\
   \bm B \bm v^T -\bm v \bm B^T
   \end{pmatrix}, \\
\bm F_{\rm nonid}  = \begin{pmatrix}
   0 \\
   0 \\
   \eta_{\rm OR} \nabla \bm B  + \eta_{\rm AD}\left[ \left(\bm J\times \bm b\right) \bm b - \bm b \left(\bm J\times \bm b\right)\right] \\
   \eta_{\rm OR} \bm J \times \bm B +\eta_{\rm AD} \left\{\left[\left(\bm J \times  \bm B \right) \times \bm b \right] \times \bm b \right\}\\
   \end{pmatrix}, 
   \label{eq:nonidelMHDSourceTerms}
\end{align}
where $\rho$, $\bm v$, $e$, $\bm B$, $\bm b = \bm B / \left|\bm B\right|$, $P$, and $\bm J = \nabla \times \bm B$ are the density, velocity, total energy per unit mass, magnetic field strength, direction of the magnetic field, pressure, and electric current, respectively.
The specific energy $e = u + \frac{1}{2} \bm v^2 + \frac{1}{2 \rho} \bm B^2$ consists of the thermal energy  per mass $u$, the kinetic energy density $\frac{1}{2} \bm v^2$, and the magnetic field energy density $\frac{1}{2 \rho} \bm B^2$. The pressure $P=p_{\rm gas} + \frac{1}{2}\bm B^2$ includes a thermal and a magnetic component. The system of equations is closed by the equation of state (EOS), which expresses $p_{\rm gas}$ as a function of the other thermodynamical quantities.
In this paper, we mostly use an isothermal EOS,
\begin{equation}
     p_{\rm gas} = \rho c_s^2,
\end{equation}
with constant isothermal sound speed $c_s$.
In some cases we also use an adiabatic equation of state
\begin{equation}
     p_{\rm gas} = \left(\gamma -1 \right) \rho u,
\end{equation}
where we introduced the adiabatic coefficient $\gamma$.
The resistivities $\eta_{\rm OR}$ and $\eta_{\rm AD}$ describe the strength of the Ohmic and ambipolar diffusion, respectively, which are in general a function of the magnetic field strength and the local chemical composition.

\subsection{Ideal MHD on a moving mesh in AREPO}

To solve equation (\ref{eq:nonidealMHDEquations}) we will use the {\small AREPO} code \citep{springel2010pur, weinberger2020arepo}, which employs a moving, unstructured Voronoi mesh in combination with the finite volume method. The general algorithms for mesh construction as well as hydrodynamic evolution were described in \cite{springel2010pur} and extended to ideal MHD in \cite{pakmor2011magnetohydrodynamics} and \cite{pakmor2013simulations}. \cite{pakmor2016improving} furthermore introduced an improved time integration and gradient estimate, and \cite{zier2022simulating} recently introduced a higher-order flux integration, which we use by default in our simulations for the ideal MHD fluxes as it significantly improves the accuracy in shear flows.

If $\nabla \cdot \bm B =0$ holds at any point in time, then one can show that this condition stays true also at later times for the system of partial differential equations (\ref{eq:nonidealMHDEquations}). However, this condition is not automatically fulfilled any longer for discretized versions of the equations, and this fact can lead to purely numerical instabilities. To reduce the influence of this error, {\small AREPO} can use the Powell scheme \citep{powell1999solution, pakmor2013simulations}  that diffuses the error away in case it arises, as well as Dedner cleaning \citep{dedner2002hyperbolic, pakmor2011magnetohydrodynamics} that additionally damps it. We typically use the first approach, with the exception of Section~\ref{sec:linearMRI} where we use the so-called shearing box approximation for which the Powell scheme is not defined so that we revert to Dedner cleaning.

We note that on static meshes the constrained transport method \citep{evans1988simulation} for divergence control is often used, which by construction guarantees $\nabla \cdot \bm B =0$ to machine precision even for the discretized equations. For non-static meshes, the implementation of this method becomes very difficult \citep{mocz2014constrained} and also tends to lead to a more diffusive behaviour.

\subsection{Implementation of non-ideal MHD in {\small AREPO}}
In this section, we discuss the implementation of the non-ideal MHD term in equation (\ref{eq:nonidealMHDEquations}), for which we apply a Strang splitting scheme which allows us to treat the non-ideal MHD flux in isolation. Our approach is originally based on the solver for cosmic ray diffusion presented in \cite{pakmor2016semi}, which was also used with slight modifications for Braginskii viscosity \citep{berlok2020braginskii}. Following the idea of the finite volume method, we can integrate the non-ideal part of equation (\ref{eq:nonidealMHDEquations}) over the volume $V_i$ of a Voronoi cell $i$.
We then find:
\begin{equation}
\frac{\partial \bm Q_i}{\partial t} = - \int_{V_i} \nabla \cdot \bm F_{\rm nonid} = - \int_{\partial V_i} \hat{n} \cdot \bm F_{\rm nonid},
\end{equation}
where we introduced the vector of conserved quantities $\bm Q_i = \int_{V_i}\bm U$ and applied Gauss's theorem. The time integral can be approximated with a second-order Runge-Kutta scheme while the surface integral can be split into a sum over the interfaces of each Voronoi cell.  We approximate the remaining integral over the interface between two cells by evaluating the flux function at the geometric centre of the face and multiplying it by the size of the face. As we have discussed in \cite{zier2022simulating} this approximation can lead to large errors in shear flows for purely hydrodynamic simulations, since if one assumes a velocity gradient within a cell, the momentum flux becomes a quadratic function in space. However, this error can be reduced by evaluating the flux function at several positions on the face. 

A closer inspection of equation (\ref{eq:nonidelMHDSourceTerms}) shows that only for ambipolar diffusion the flux can become a quadratic function in space (if we assume a constant electric current and a linear magnetic field). We therefore only observe a marginally improved accuracy (around 1\%) for the test problems in this paper if we apply this higher-order flux integration also for the non-ideal MHD terms. We, therefore, do only use the higher order flux integration for the ideal MHD term. We are then left with the problem of how to calculate the flux function at the interface.

\subsubsection{Gradient estimation at an interface}
\label{subsubsec:estimateGradients}

A closer look at equation~(\ref{eq:nonidelMHDSourceTerms}) shows that we need accurate estimates of the magnetic field as well as its gradients at cell interfaces. While the Ohmic diffusion term only requires the normal component of the gradient of the magnetic field at each interface, the ambipolar diffusion term also requires the components parallel to the surface. \cite{pakmor2016semi} introduced in their Section~2.1 a method to obtain those gradients by performing a least-squares fit (LSF) at the corners of the Voronoi mesh using the values at the three (four) adjacent cell centres in 2D (3D). The gradients as well as the magnetic field at the center of the face can then be defined as a weighted sum of the corresponding values at the corners. While testing this method for this paper we found that the LSF at the corners can sometimes become unstable, prompting us to find a fallback strategy to still obtain the gradients if this happens.  In this case we initially tried a finite difference approach to calculate the normal component of the gradient, setting the the other components to zero. But this approach became unstable for magnetic shocks, motivating a more profound change of our scheme.

Instead of combining the results of the LSFs at the corners of a Voronoi cell, we now collect all cells that contribute to one corner of the interface and then perform one large, over-determined LSF at the centre of the interface. If the interface has $K$ corners, we obtain $M=K+2$ cells as input (the two cells forming the interface contribute to all corners, each other cell to two corners). The residuals after the fit are given by
\begin{equation}
 r_i = \phi\left(\bm c\right) + \left(\nabla \phi \right) \left(\bm s_i - \bm c \right) - \phi \left(\bm s_i \right),    
\end{equation}
where $\phi$ is the quantity of interest (in our case the three components of the magnetic field), $\phi\left(\bm c\right)$ its unknown value at the center of the interface, $\nabla \phi$ its gradient at the interface, $\bm s_i$ the centre of mass of cell $i$, and $\phi \left(\bm s_i \right)$ is the known value of $\phi$ at position $\bm s_i$. This can be rewritten as a matrix equation:
\begin{equation}
    \bm r = \bm X \bm q - \bm Y,
    \label{eq:matrixEquationResidual}
\end{equation}
where $\bm r$ and $\bm Y$ are M-vectors, $\bm q$ is an N-vector and $\bm X$ is a $M\times N$ matrix with $N=3$ ($N=4$) in 2D (3D).
The components are given by
\begin{align}
    \bm q_0 &= \phi \left(\bm c \right),\\
    \bm q_{1...N-1} &= \nabla \phi_{0...N-2},\\
    \bm Y_i &= \phi\left(\bm s_i\right),\\
    \bm X_{i,0} &= 1,\\
    \bm X_{i,1...N-1} &= \bm s_{i,0...N-2} - \bm c_{0...N-2}.
    \label{eq:componentOfX}
\end{align}
Our goal is now to minimize the residual and therefore find the so-called pseudo-inverse $\bm X^{-1}$. In this case the required vector $\bm q$ is just given by a simple matrix multiplication:
\begin{equation}
    \bm q = \bm X^{-1} \bm Y.
    \label{eq:invertedLSF}
\end{equation}
In Appendix~\ref{app:leastSquareFit}, we discuss the details of how we obtain $\bm X^{-1}$ and also the numerical stability of this procedure. We also introduce a weighting for each cell, for which we first assign a weight $w_i$ to each corner, as in \cite{pakmor2016semi}.  This weight is equally split between the 3 (4) adjacent cells. Finally, we perform a sum over all corners to determine the full weight of each cell. A closer inspection of equation (\ref{eq:invertedLSF}) shows that the magnetic field at the interface is given by a weighted sum of the magnetic fields at the centre of mass of the Voronoi cells.  The weights are given by the first row of matrix $\bm X^{-1}$, which can be positive or negative but their sum has to be equal to $1$.

Note that our finite volume scheme can become unstable if we create new extrema by the linear interpolation. This is potentially the case if the sum of all negative entries in the first row becomes too small, and we therefore use this sum as a measure of the stability of the LSF. If the sum is smaller than $-0.2$, we typically reject the results from this method and instead use as a fallback a finite difference scheme we discuss below. In the simulations presented in this paper the fallback only has to be used for fewer than 0.5\% of all interfaces. Interestingly, the number of interfaces below a certain threshold decreases significantly faster than for a LSF at the corners.

As a fallback,  we calculate only the normal component of the gradient by:
\begin{equation}
\nabla \phi_{\rm face, n} = \frac{\phi_L - \phi_R}{\left| \bm c_L - \bm c_R \right|}
\left(\frac{\bm c_L - \bm c_R}{\left| \bm c_L - \bm c_R \right|} \right) \cdot \hat{n}_{\rm face},
\label{eq:fallBack}
\end{equation}
where $L$ and $R$ denote the left and right cells of the face. We set the value of the magnetic field to the arithmetic average of the values of the two neighbouring cells and the component of the gradient parallel to the face to zero. This mostly becomes important for ambipolar diffusion, since for Ohmic diffusion we only require the normal component.

\subsubsection{Time step constraints}

The diffusion terms add additional constraints to the largest allowed time step for a stable integration. Following \cite{bai2014hall} we can write the time step constraint as:
\begin{equation}
\Delta t_{\rm nonideal} = C_{\rm nonideal}\frac{\Delta x_i^2}{4 d\,\eta_{\rm tot}}, \label{eq:timeStepConstrainedDiffusion}
\end{equation}
where $\Delta x_i = \left[ 3V_i / \left(4 \pi\right) \right]^{1/3}$ is the effective radius  of the Voronoi cell $i$ with volume $V_i$ in three dimensions ($\sqrt{V_i /\pi}$ in two dimensions), $d$ the number of dimensions, and $\eta_{\rm tot} = \eta_{\rm OR} + \eta_{\rm AD}$ the total diffusivity. We also introduced the non-ideal Courant number $C_{\rm nonideal}$ which should be smaller than 1 to assure stability. We typically set it to $0.5$. We use local time steps in this paper with the timestep hierarchy presented in \cite{weinberger2020arepo} and set the maximum time step of each cell to
\begin{equation}
\Delta t_i = {\rm min} \left(\Delta t_{\rm ideal}, \Delta t_{\rm nonid}\right).
\end{equation}

As one can see in equation~(\ref{eq:timeStepConstrainedDiffusion}), the diffusion time steps are indirectly proportional to the square of the spatial resolution, which means that in dense regions (small cells) the time step constraint can become prohibitively expensive. To mitigate this,  \cite{berlok2020braginskii} implemented a second-order accurate super-time stepping scheme for Braginksii viscosity, which can significantly speed up the code when diffusion effects begin to dominate the overall time step constraint. However, this approach can become unstable for local time steps (private communication with Rüdiger Pakmor) and can therefore only be safely used with global time steps. Global time steps can become prohibitively expensive in simulations with a deep time step hierarchy. For this reason, we do not use super-time stepping with global time steps in this paper.

\subsubsection{The strength of non-ideal diffusivities}

\label{subsubsec:strengthDiffusivities}

We have parameterized the strength of Ohmic and ambipolar diffusion using the diffusivities $\eta_{\rm OR}$ and $\eta_{\rm AD}$. These coefficients are highly dependent on the local chemical composition and magnetic field structure. Although specific computational methods can be used to include these coefficients self-consistently \citep[e.g.][]{wurster2016nicil}, the results are still affected by unknown factors such as the cosmic ray ionization rate, or the grain-size distribution of dust. Since our main objective in this paper is to demonstrate the accuracy of our implementation of non-ideal MHD, we will therefore only use constant or parameterized values for the diffusivities. 

The diffusion coefficient for ambipolar diffusion can be expressed as:
\begin{equation}
    \eta_{\rm AD} = \frac{B^2}{\gamma_{\rm AD}\rho\rho_i},
\end{equation}
where $\gamma_{\rm AD}$ represents the drag coefficient between ions and neutrals, $\rho$ is the density of neutrals, and $\rho_i$ is the ion density \citep{choi2009explicit}. The ion density itself can be modelled as a power law \citep[e.g.][]{elmegreen1979magnetic}:
\begin{equation}
    \rho_i = \rho_{i0} \left(\frac{\rho}{\rho_0} \right)^\alpha,
\end{equation}
where we will use $\alpha = 0$ as a simplification in this paper. In simulations of rotating disks, the strength of the non-ideal effects can also be described by the Elsasser numbers in terms of the ratio of the magnetic and Coriolis forces:
\begin{equation}
    \Lambda_{\rm OR, AD} = \frac{v_A^2}{\Omega \left|\eta_{\rm OR, AD} \right|},
    \label{eq:elssaserNumber}
\end{equation}
where $\Omega$ is the local rotational frequency, and $v_A =B /\sqrt{\rho}$ is the Alfvén velocity \citep{wurster2016nicil}.

\begin{table}
    \centering
    \begin{tabular}{c|c|c|c}
    \hline
        Effect & Dimension & $t_0$ & $B_{\rm con, c}$ \\
        \hline
 Ohmic & 1D& 0.01 & 1\\
 Ohmic & 2D& 0.05 & 1\\
 Ambipolar & 1D& 0.001 & 0.15\\
 Ambipolar & 2D& 0.0001& 0.078535\\
 \hline
    \end{tabular}
    \caption{Details for the initial conditions for the magnetic diffusion tests. We initialize the simulation at time $t = t_0$. The conserved quantity $B_{\rm con, c} = \int_{\bm x} B_y\, {\rm d}x$ characterizes the initial amplitude.}
    \label{tab:DetailsDiffusion}
\end{table}
 
 \begin{figure}
    \centering
    \includegraphics[width=1\linewidth]{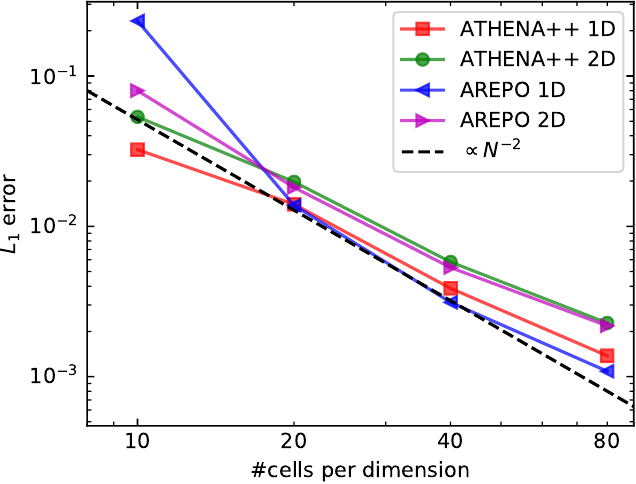}
    \caption{The $L_1$ norm of the error of $B_y$ as a function of resolution for the pure diffusion tests with Ohmic diffusion, as discussed in Section~\ref{subsubsec:ohmicDiffusion}.}
    \label{fig:L1_error_diffusion_ohmic}
\end{figure}

\subsection{The comparison code {\small ATHENA++}}

To better understand the advantages and disadvantages of our new non-ideal MHD solver on a moving mesh we repeat all the test simulations presented in this paper with the publicly available Eulerian code {\small ATHENA++} \citep{stone2020athena++}. The code is the successor of the {\small ATHENA} code which has frequently been used for non-ideal MHD studies, making it an ideal benchmark for comparisons. In our simulations with {\small ATHENA++}, we use its second-order finite volume mode with linear reconstruction, a second-order accurate time integration scheme, and the HLLD Riemann solver. These are characteristics similar to the methods used in {\small AREPO}. However, unlike {\small AREPO}, {\small ATHENA++} uses a structured mesh that is stationary, and it employs constrained transport \citep{evans1988simulation} instead of a divergence cleaning method to evolve the magnetic field. This ensures that $\nabla \cdot B = 0$ holds up to machine precision in {\small ATHENA++}.

\section{Code tests and convergence}
\label{sec:Tests}

\begin{figure}
    \centering
    \includegraphics[width=1\linewidth]{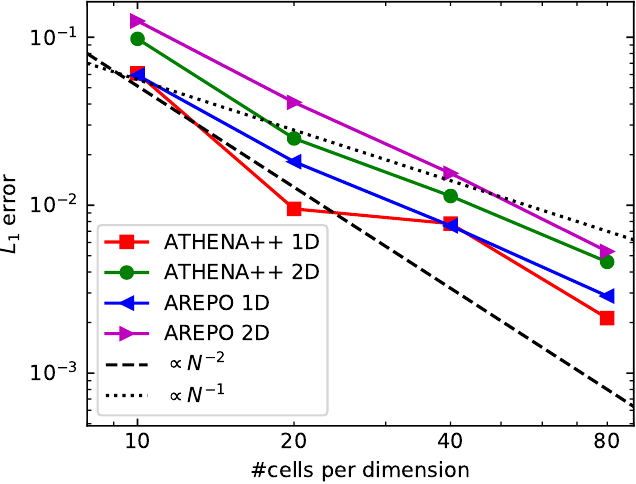}
    \caption{The $L_1$ norm of the error of $B_y$ as a function of resolution for the pure diffusion tests with ambipolar diffusion, discussed in Section~\ref{subsubsec:ambipolarDiffusion}.}
    \label{fig:L1_error_diffusion_ambipolar}
\end{figure}

\begin{figure}
    \centering
    \includegraphics[width=1\linewidth]{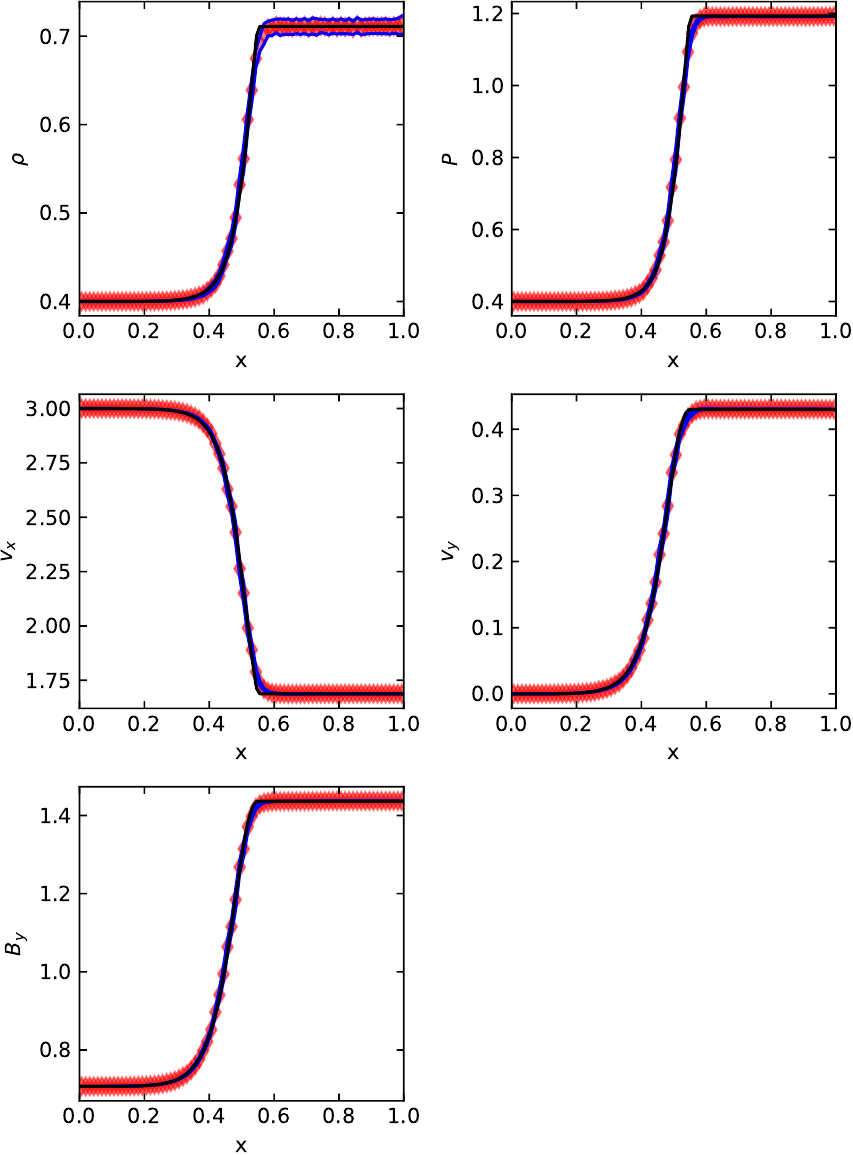}
    \caption{The structure of a non-isothermal, oblique C-shock with Ohmic diffusion. The black line corresponds to the semi-analytically expected structure of the shock, while the red symbols show the average values in a simulation with {\small AREPO} at time $t=5$, with on average 40 cells per dimension (we use 80 bins in the $x$-direction for this plot). The blue lines show the  $\pm 1$ standard deviation in each bin. More details can be found in Section~\ref{subsubsec:CShockOhmic}.}
    \label{fig:OhmicNonIsothermal_t5_40_cells}
\end{figure}

\begin{table*}
    \centering
    \begin{tabular}{c|c|c|c|c|c|c}
    \hline
        Variable & $\rho$ & $v_x$ & $v_y$ & $B_x$ & $B_y$ & $P$\\
        \hline
        Ohmic: Pre-shock values & 0.4 & 3& 0& $\sqrt{2}$/2 & $\sqrt{2}$/2 &0.4\\
        Ohmic: Post-shock values (isothermal) & 0.85489 &	1.40369 &	0.67328 & $\sqrt{2} /2$ & 1.84969 &0.85489\\
        Ohmic: Post-shock values (adiabatic) & 0.71084 &	1.68814 &	0.4299 & $\sqrt{2} /2$ & 1.43667 &1.19222\\
        Ambipolar: Pre-shock values & 0.5 & 5& 0& $\sqrt{2}$ & $\sqrt{2}$ &0.125\\
         Ambipolar: Post-shock values (isothermal) & 1.0727	&2.3305 &1.3953 & $\sqrt{2}$ & 3.8809 	&0.2681\\
         Ambipolar: Post-shock values (adiabatic) & 0.9880 &	2.5303 & 1.1415 & $\sqrt{2}$ & 3.4327& 1.4075\\
        \hline
    \end{tabular}
    \caption{Initial conditions for the C-shock tests discussed in Section~\ref{subsec:CShock}.
    For the isothermal solutions, we use $c_s = 1$ in the Ohmic diffusion case and $c_s = 0.5$ for the ambipolar diffusion case.}
    \label{tab:CShockInitialConditions}
\end{table*}

In this section, we test our implementation of Ohmic and ambipolar diffusion in {\small AREPO} using standardised test problems that allow an analytical solution to compare our results with. For a realistic comparison we first construct an unstructured mesh by simulating the ground state of the shearing box as presented in \cite{zier2022simulating}. We perform these simulations in 3D with initially 10 cells per dimension. We use the resulting mesh as the initial condition and generate larger meshes by combining the initial small meshes. The cells are then initialised with the value of the continuous initial conditions at their centre of mass. For the simulations with {\small ATHENA++} we use a static, uniform Cartesian grid with the same number of cells as in {\small AREPO}. To measure the error in our results we define the $L_1$ error by:
\begin{equation}
L_1 = \frac{\frac{1}{V} \sum_{i =1}^{N_{\rm cell}}\left| f_i\right|}{\frac{1}{V} \sum_{i =1}^{N_{\rm cell}}\left|X_t\right|},
\end{equation}
where $\left| f_i\right|$ is the difference between the value of cell $i$ and the theoretical value of the quantity $X_t$ at its centre of mass. The sum is generally over all cells in the simulation. In Appendix~\ref{app:analyticalSolutions} we provide the analytical solutions for all tests discussed in this section.

\subsection{Diffusion of a magnetic field}

We first test the diffusion of a magnetic field using the non-ideal MHD terms similar to \cite{masson2012incorporating}. This means that we do not follow the hydrodynamical evolution of the gas and keep the grid static. 
We will assume $B_x = B_z = 0$, which together with the $\nabla \cdot \bm B =0$ condition leads to $B_y = B_y(x,z)$.
This allows two different types of initial conditions for four tests:
\begin{align}
    B_{\rm y, 1D} = B_{y0}\delta(x),
    \label{eq:ICDiffusion1D}\\
    B_{\rm y, 2D} = B_{y0} \delta (x) \delta (z),
     \label{eq:ICDiffusion2D}
\end{align}
where we introduced the $\delta$-distribution. In this situation we end up with purely diffusive differential equations, with the difference being that in the ambipolar case the diffusion constant depends on the magnetic field strength.  In all simulations we use a box of size $L_x \times L_y \times L_z = 1 \times 1 \times 1$ with centre $(0,0,0)$. 

To initialise our simulations, we set up the analytical solutions presented in Appendix~\ref{app:solutionsDiffusion} at a finite time $t_0$ and run the simulations until time $t = 1$. The quantity $B_{\rm con} = \int_{\bm x} B_y\,{\rm d}x$ is conserved in this test and is therefore a measure of the amplitude of the initial peak. Because we sample the analytical solution with a finite number of cells at the beginning, the total $B_{\rm con, s}$ may not exactly agree with the result using the continuous field $B_{\rm con, c}$. This can lead to systematic errors, so we first sample the solution at $t = t_0$, measure $B_{\rm con,s}$, and then multiply $B_y$ by $B_{\rm con, c} / B_{\rm con, s}$.  In \cref{tab:DetailsDiffusion}, we list $t_0$ as well as $B_{\rm con, c}$ for our simulations. Unlike in the other sections, we will always use the result of the least squares fit. The reason is that a static mesh is only used in this particular test for the {\small AREPO} simulations, as otherwise errors due to the approximation (\ref{eq:fallBack}) would build up at the same position in the simulation box.

\subsubsection{Ohmic diffusion}
\label{subsubsec:ohmicDiffusion}

Assuming a constant Ohmic diffusivity $\eta_{\rm OR}$ we find the standard diffusion equation:
\begin{equation}
    \frac{\partial B_y}{\partial t} = \eta_{\rm OR} \Delta  B_y,
    \label{eq:diffusionEquationOhmic}
\end{equation}
which has the solutions (\ref{eq:solutionDiffusionOhmic1D}) and (\ref{eq:solutionDiffusionOhmic2D}). We use $\eta_{\rm OR} = 0.01$ and show in \cref{fig:L1_error_diffusion_ohmic} the results of our resolution study. We find, as expected, an approximate second order convergence for both codes and setups. Except for the lowest resolution simulations, where we also had to significantly modify the initial condition to get the correct results for $B_{\rm con, s}$, the errors are very similar for {\small AREPO} and {\small ATHENA++}.

\begin{figure}
    \centering
    \includegraphics[width=1\linewidth]{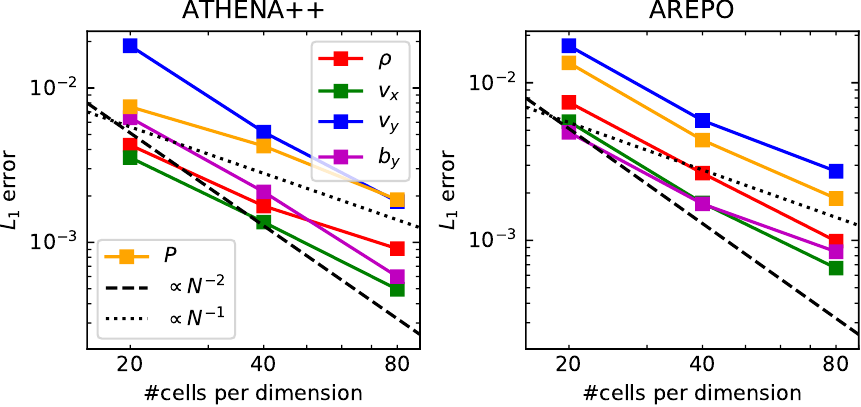}
    \caption{The $L_1$ error norm as a function of resolution for the non-isothermal C-shock test with Ohmic diffusion. We show results obtained with {\small ATHENA++} (left) and {\small AREPO} (right).
    More details can be found in Section~\ref{subsubsec:CShockOhmic}.}
    \label{fig:L1_error_CShock_Ohmic_NonIsothermal}
\end{figure}

\begin{figure}
    \centering
    \includegraphics[width=1\linewidth]{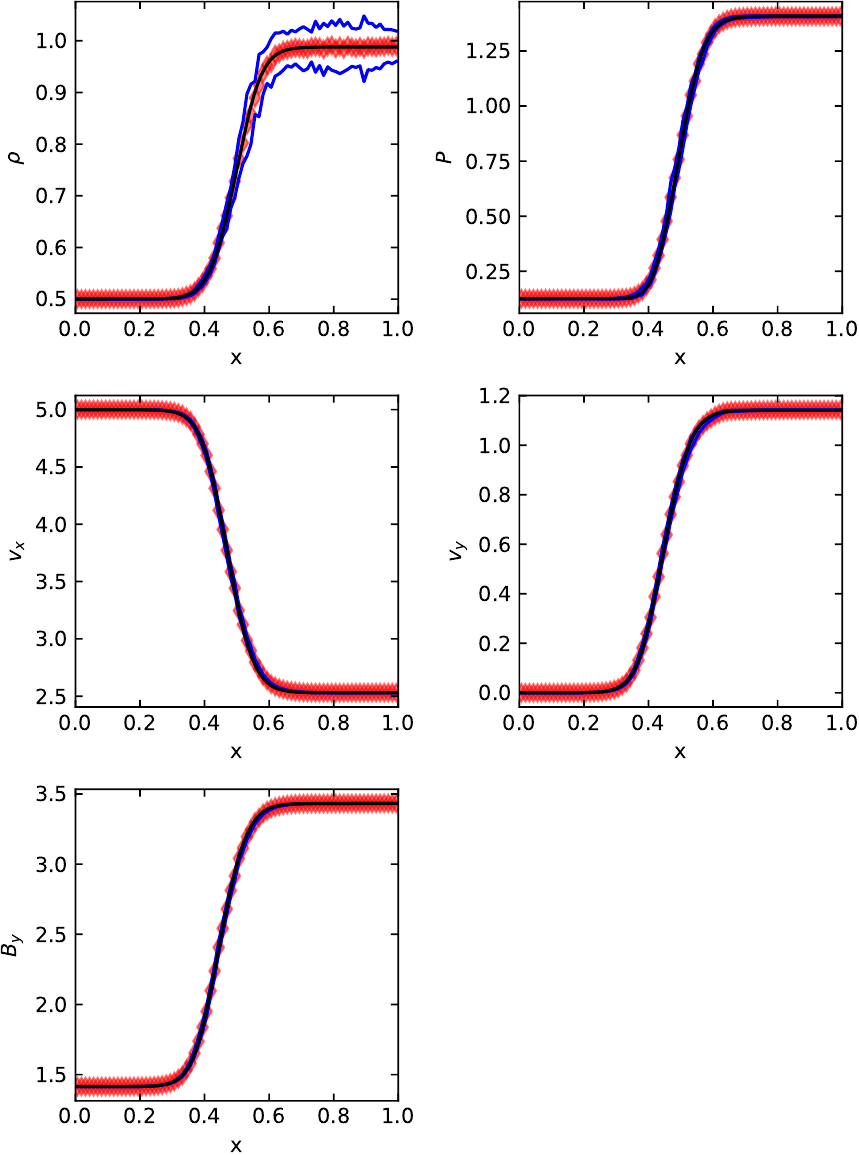}
        \caption{The structure of a non-isothermal, oblique C-shock with ambipolar diffusion. The black line corresponds to the semi-analytically expected structure of the shock, while the red symbols show the average values in a simulation with {\small AREPO} at time $t=5$, with on average 40 cells per dimension (we use 80 bins in the $x$-direction for this plot). The blue lines show the  $\pm 1$ standard deviation in each bin. More details can be found in Section~\ref{subsubsec:CShockAmbipolar}.}
    \label{fig:AmbipolarNonIsothermal_t5_40_Cells}
\end{figure}

\begin{figure}
    \centering
    \includegraphics[width=1\linewidth]{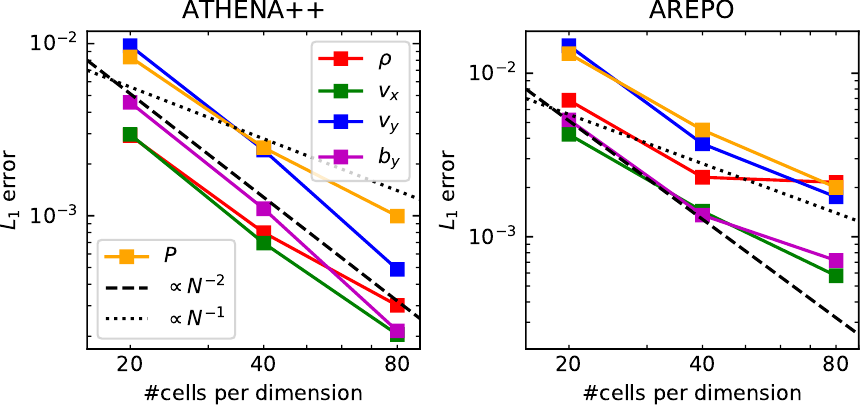}
    \caption{The $L_1$ error as a function of resolution for the non-isothermal C-shock test with ambipolar diffusion. We show results obtained with {\small ATHENA++} (left) and {\small AREPO} (right).
    More details can be found in Section~\ref{subsubsec:CShockAmbipolar}.}
    \label{fig:L1_error_CShock_Ambipolar_NonIsothermal}
\end{figure}

\subsubsection{Ambipolar diffusion}
\label{subsubsec:ambipolarDiffusion}

For ambipolar diffusion, we aim to solve the diffusion equation
\begin{equation}
    \frac{\partial B_y}{\partial t} = \nabla \cdot \left(\frac{v_A^2}{\gamma_{\rm AD } \rho_i} \nabla B_y\right)
\end{equation}
with $v_A = B_y / \sqrt{\rho}$. The solution is given by the Barenblatt-Platte solution presented in  Appendix~\ref{app:barenBlattSolution}, and we choose $\gamma_{\rm AD} \rho_i \rho = 1$. As can be seen in \cref{fig:L1_error_diffusion_ambipolar}, the errors for both codes are very similar, though slightly smaller for {\small ATHENA++}. The convergence rate lies between second and first order in both cases, but seems to be slightly better for {\small AREPO}, which is reflected in a decreasing difference in the errors between the codes.

\subsection{C-shock}
\label{subsec:CShock}

While magnetic shocks in the ideal MHD regime contain discontinuous jumps (J-type shock), the non-ideal MHD effects make these shock features continuous \citep[C-type shock,][]{Draine1980}. The exact structure of the shock depends on the type of diffusion as well as on its strength, making this problem a popular test for implementations of non-ideal MHD \citep[e.g.][]{MacLow1995, Duffin2008, Bai2011, masson2012incorporating}. In this section, we repeat the tests presented in \cite{masson2012incorporating} which cover the effect of Ohmic and ambipolar diffusion as well as an isothermal and adiabatic equation of state. In \cref{tab:CShockInitialConditions} we present the pre-shock and post-shock values for all tests where we choose the $x$-direction as the shock direction and set $B_z = v_z = 0$. In this section, we will only discuss the results for an adiabatic equation of state, as this test also covers the energy flux between cells. As the isothermal tests are very popular in the literature, we present our results for them in the Appendix~\ref{app:resultsIsothermalCShock}. We also specify in  Appendix~\ref{app:CShockEquations} the equations we solve to obtain the detailed structure of the C-type shock, which we use as a benchmark for our simulations. 

For our tests we choose a box of size $L_x \times L_y \times L_z = 1 \times 1 \times 1 $ and as initial conditions we set the pre-shock values for $x < 0.5$ and the post-shock values for $x> 0.5$. In the $y$- and $z$-directions, we use periodic boundary conditions while in the $x$-direction for {\small ATHENA++} we set the primitive variables of the ghost cells to the pre- and post-shock values, respectively. In the simulations with {\small AREPO} the cells move through the shock, which means that we have to continuously create new cells at $x =0$ and remove cells at $x=1$. We, therefore, use periodic boundary conditions also in the $x$-direction, but we extend the box so it spans the interval $-0.5 < x<1.5$. Between $-0.5 < x < 0$ the properties of cells are continuously set to the pre-shock values while for $1 < x < 1.5$ we set them to the post-shock values. This setup also keeps the number of cells constant, which makes it easier to compare it with the simulations run with {\small ATHENA++}. The position of the shock can move for both codes, so we fit the semi-analytic profiles to the simulation results by shifting the former in the $x$-direction and minimising the $L_1$ error for the density. The moving mesh in {\small AREPO} is unstructured and therefore the hydrodynamic quantities can also be functions of the $y-$ and $z-$ coordinates. This can lead to noise in the post-shock region, which can be observed even for purely hydrodynamic shocks and is therefore not caused by inaccuracies of the non-ideal MHD solver. We therefore bin all quantities in the $x$-direction for the {\small AREPO} simulations and calculate the mean as well as the standard deviation in each bin. To fit the semi-analytical profiles to the simulation results we use the averaged density profiles and use twice the number of bins as we have cells in the $x$-direction in the corresponding  {\small ATHENA++} simulation.

\begin{figure*}
    \centering
    \includegraphics[width=1\linewidth]{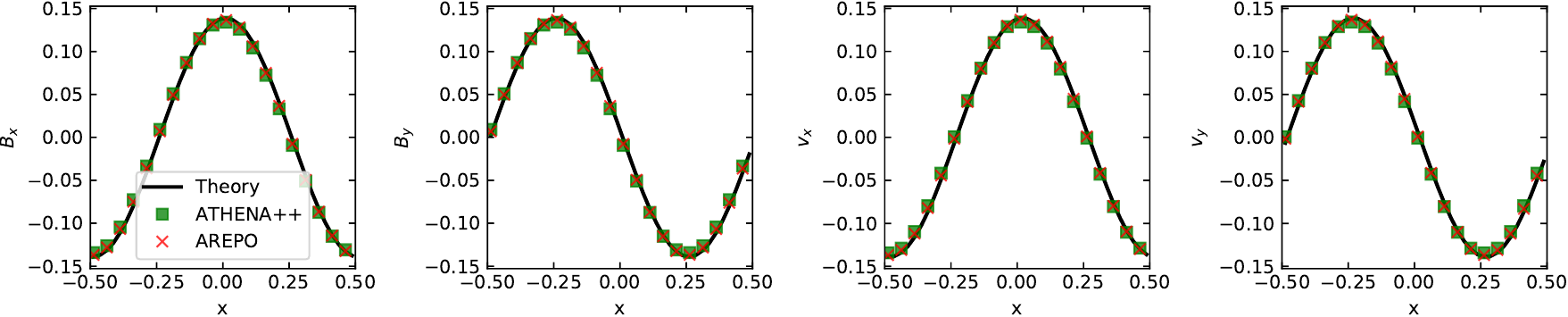}
    \caption{The damping of a travelling Alfvén wave by Ohmic diffusion. We show the structure at time $t=5$ for simulations with {\small ATHENA++} (green) and {\small AREPO} (red), with an effective resolution of 40 cells per wave length and compare the results with the analytic solution (black line). For better visibility we first bin the results into 40 bins in the $x$-direction and only show the average in every second bin.}
    \label{fig:OhmicTravelling_structure}
\end{figure*}

\begin{figure}
    \centering
    \includegraphics[width=1\linewidth]{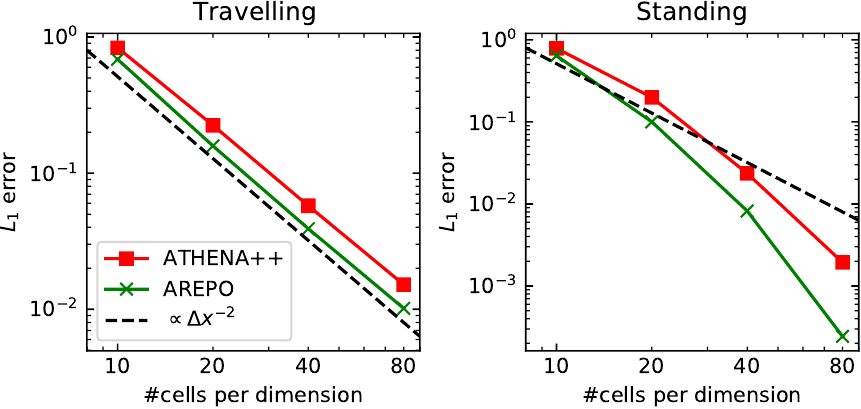}
    \caption{The $L_1$ error norm of the magnetic field component $B_y$ as a function of resolution for simulations of the damping of an Alfvén wave by Ohmic diffusion. We measure the errors at $t=5$ and show on the left side the results for a travelling wave, and on the right side for a standing wave.}
    \label{fig:L1_error_Ohmic_Alvfen}
\end{figure}

\subsubsection{Ohmic diffusion}
\label{subsubsec:CShockOhmic}

As we show in \cref{fig:OhmicNonIsothermal_t5_40_cells}, our implementation in {\small AREPO} is able to accurately model the structure of the shock with an Ohmic diffusion strength of $\eta_{\rm OR} = 0.1$. Only in the density profile we see a significant amount of noise in the post-shock region, which we would also observe without non-ideal MHD. The pressure profile displays significantly less noise than the density profile, which means that overdense regions correspond to colder than expected regions. The approximately constant pressure therefore allows the density perturbations  to be more or less passively advected in the post-shock region. This is in contrast to the isothermal case as discussed in Appendix~\ref{app:resultsIsothermalCShock}, where density perturbations always lead to pressure perturbations, which get washed out. In \cref{fig:L1_error_CShock_Ohmic_NonIsothermal} we can see that both codes achieve between first and second order convergence, but in general the errors are larger for {\small AREPO}. This can be explained by the post-shock noise in the density, which does not diminish as we increase the resolution and can also indirectly affect the other quantities of interest. We also note that the fitting of the semi-analytic profiles to the results of the simulations can lead to small errors, since we only try to minimise the error in the density and not in the other quantities. As expected, we also observe close to second order convergence for the isothermal case.

\begin{figure}
    \centering
    \includegraphics[width=1\linewidth]{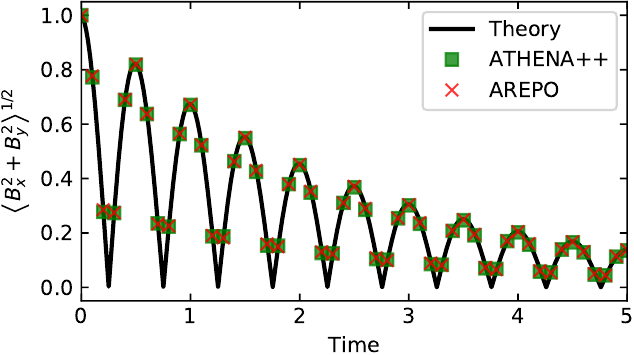}
    \caption{The evolution of the rms magnetic field for a standing Alfvén wave damped by Ohmic diffusion. We show the results for simulations with 40 cell per dimension and two different codes.}
    \label{fig:OhmicStandingB2}
\end{figure}

\subsubsection{Ambipolar diffusion}
\label{subsubsec:CShockAmbipolar}

For the ambipolar diffusion simulations we choose $\rho_i = 1$ and $\gamma_{\rm AD} = 75$ and show in \cref{fig:AmbipolarNonIsothermal_t5_40_Cells} that {\small AREPO} can resolve the expected shock structure. As before, we observe noise in the post-shock regions that is larger than in the case of ohmic diffusion, which can be explained by the stronger shock. It can also be observed in an ideal MHD shock with the same setup, and again it can be passively advected in the post-shock region due to the approximately constant pressure. For an isothermal equation of state, density perturbations however lead to pressure perturbations, which is why the noise is much smaller in this case. In \cref{fig:L1_error_CShock_Ambipolar_NonIsothermal} we show the results of a convergence study and find for {\small ATHENA++} an approximate second order convergence. The absolute errors are larger for {\small AREPO}, and we are able to establish second order convergence only for low resolution. In particular, the average density error does not decrease when we change from 40 to 80 cells per dimension, which also affects the other quantities. This can again be attributed to the density noise, and is not due to the non-ideal MHD solver itself.

\subsection{Damping of an Alfvén wave by Ohmic diffusion}
\label{subsec:dampingAlfvenWave}

In order to study the interaction between the magnetic field and the velocity field, in this section we consider the attenuation of both, a standing and a travelling Alfvén wave. For small perturbations, the equations for ambipolar diffusion simplify to those for a linear Alfvén wave with Ohmic diffusion, so we will concentrate on the latter case. Following \cite{masson2012incorporating}, we set up a ground state with $\rho_0 = 1$, $\bm v = 0$ and $\bm B = \left(0,0, B_z \right)$, which leads to a dispersion relation with the solution:
\begin{equation}
    s = -\frac{\eta_{\rm OR} k^2}{2} \pm i \sqrt{k^2 v_A^2 - \left(\frac{\eta_{\rm OR} k^2}{2} \right)^2}.
    \label{eq:eigenvalueDampingAlfven}
\end{equation}
Here we have assumed a perturbation with wave vector $\bm k = k \hat{e}_z$, where $v_A = B_z /\sqrt{\rho_0}$ is the Alfvén velocity.
If the term inside the square root is positive, the system performs damped oscillations, the detailed structure of which is described in Appendix~\ref{app:dampedAlfvenWave}. For our tests we set up a periodic box of size $L_x = L_y = L_z = 1$, with magnetic field $B_z = 1$, density $\rho_0 = 1$, diffusivity $\eta_{\rm OR} = 0.02$, and wave vector $k_z = 2\pi$. We run all simulations until we reach a time $t = 5$, using an amplitude of $\delta B = 1$ for the initial perturbation.

\subsubsection{Travelling wave}

The accuracy of both codes in modeling the evolution of a travelling wave and preserving the structure of the initial wave is shown in \cref{fig:OhmicTravelling_structure}. The results demonstrate that both codes are able to accurately capture the evolution of the travelling wave while maintaining the integrity of its original structure. In \cref{fig:L1_error_Ohmic_Alvfen} we present the results of a resolution study in which we find for travelling waves, for both velocity and magnetic field components, the same relative errors. Additionally, we observe perfect second-order convergence in both codes. The absolute values of the errors are comparable as well, although slightly smaller in the case of {\small AREPO}.

\subsubsection{Standing wave}

In the case of a standing wave, the amplitudes of the magnetic field and the velocity undergo oscillations. This means that periodically the perturbed magnetic field completely disappears. \cref{fig:OhmicStandingB2} illustrates these oscillations, and both codes accurately describe them. The damping rate and frequency of the oscillations are also well captured. Calculating the relative $L_1$ error for the average magnetic field strength can be more complex since it does not evolve monotonically over time. It may involve errors in the oscillation frequency, damping rate, and modifications to the wave structure. Nonetheless, in \cref{fig:L1_error_Ohmic_Alvfen}, we present the results of a convergence study that demonstrates second-order or even faster convergence rate for the magnetic field in both codes.

\begin{figure}
    \centering
    \includegraphics[width=1\linewidth]{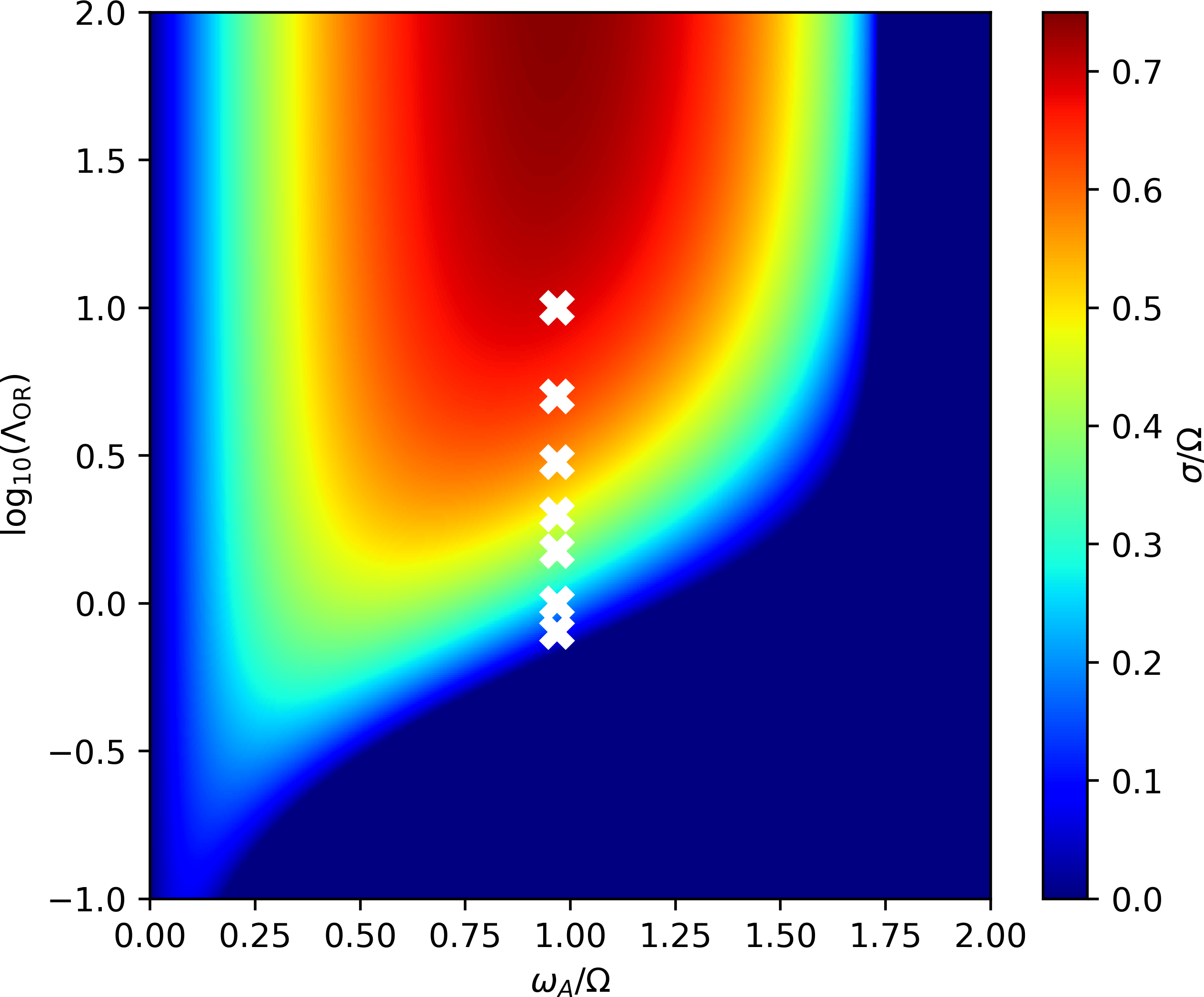}
    \caption{The linear growth rate of the MRI as a function of the Ohmic Elsasser number $\Lambda_{\rm OR}$ and the Alfvén frequency $\omega_A$ for an axisymmetric perturbation with $k_x = 0$. The growth rates are in this case independent of the background density and the isothermal sound speed. The white crosses correspond to the parameters for which we perform simulations in the linear regime. }
    \label{fig:linear_growth_ohmic}
\end{figure}

\begin{figure*}
    \centering
    \includegraphics[width=1\linewidth]{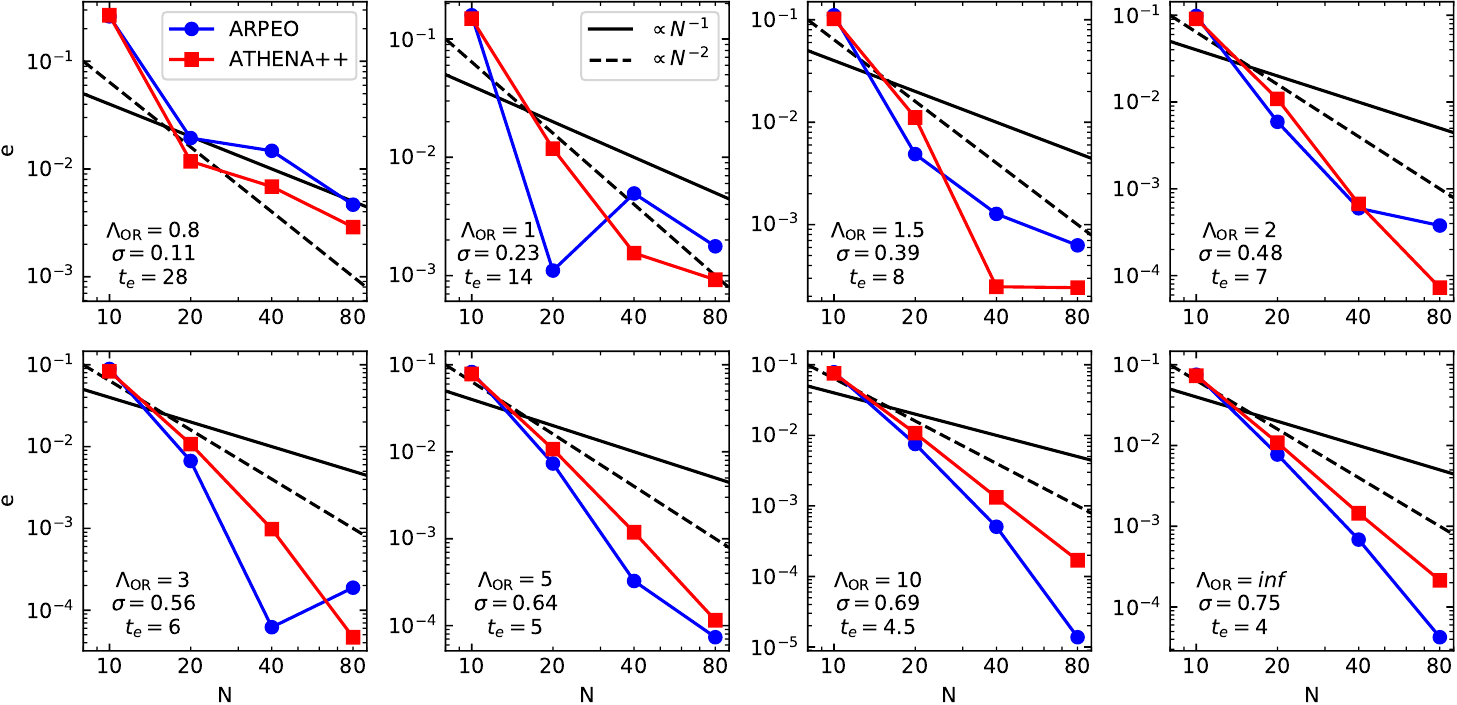}
    \caption{The relative error $e = (\sigma_{\rm theory} - \sigma_{\rm sim}) /\sigma_{\rm theory}$ of the growth rate of the MRI with Ohmic diffusion as a function of the numbers of cells $N$ per dimension for different Elsasser numbers $\Lambda_{\rm OR}$ at time $t_e$.}
    \label{fig:linear_growth_ohmic_results}
\end{figure*}

\begin{figure}
    \centering
    \includegraphics[width=1\linewidth]{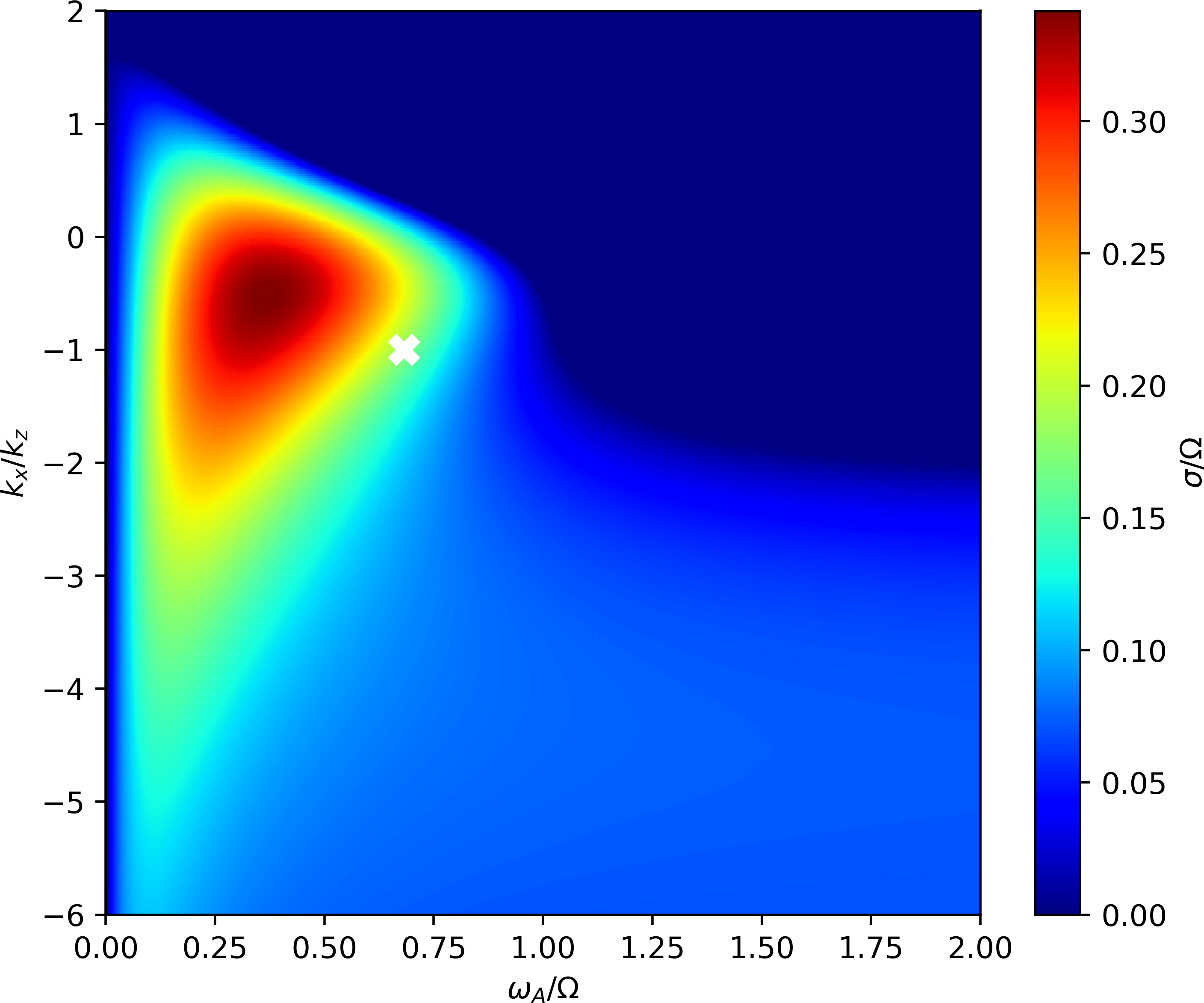}
    \caption{The linear growth rate of the MRI as a function of the radial wave number $k_x$ and the Alfvén frequency $\omega_A$ for an external magnetic field $B_{0,y} = B_{0,z}$ and Elsasser number $\Lambda_{\rm AD} = 0.8$. The white cross corresponds to the parameters we use in one of our simulations.}
    \label{fig:linear_growth_kx_ambipolar}
\end{figure}

\begin{figure*}
    \centering
    \includegraphics[width=1\linewidth]{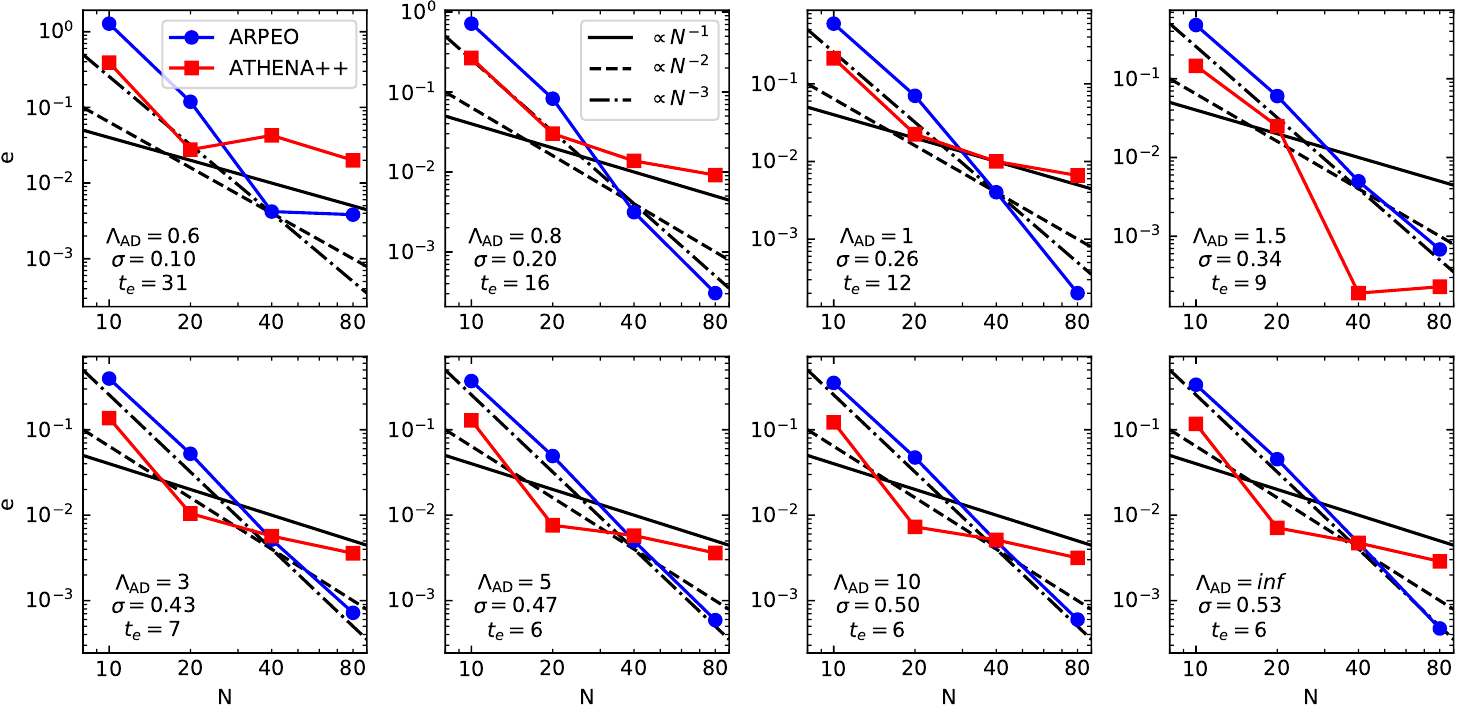}
    \caption{The relative error $e = (\sigma_{\rm theory} - \sigma_{\rm sim}) /\sigma_{\rm theory}$ of the growth rate of the MRI with ambipolar diffusion as a function of the number of cells $N$ per dimension for different Elsasser numbers $\Lambda_{\rm AD}$ at time $t_e$.}
    \label{fig:linear_growth_ambipolar_results}
\end{figure*}

\section{Linear growth of the Magneorotational instability}
\label{sec:linearMRI}

As a first more complicated test we analyse the linear growth of the magnetorotational instability with Ohmic and ambipolar diffusion. We will use the shearing box approximation we introduce in the next section combined with the last section's method to construct the initial Voronoi mesh.

\subsection{The shearing box approximation}

The shearing box approximation is widely used in studies of the magnetorotational instability since it allows a high spatial resolution and easily defined boundary conditions in contrast to global disk simulations. We solve the MHD equations in a small, rectangular box that rotates at a radius $r_0$ with the local orbital frequency $\Omega_0 = \Omega \left(r_0\right)$. We define a Cartesian coordinate system with $\hat{e}_x$ pointing in the radial direction, $\hat{e}_y$ pointing in the azimuthal direction and $\hat{e}_z$ being perpendicular to the other two vectors. The shearing box equations can be derived by adding a centrifugal force to equation (\ref{eq:nonidealMHDEquations}), and by performing a coordinate transformation into the rotating system:
\begin{equation}
      \frac{\partial \bm U }{\partial t}+ \nabla \cdot \bm F_{\rm ideal}(\bm U) +  \nabla \cdot \bm F_{\rm nonid}(\bm U)= \bm S_{\rm grav} + \bm S_{\rm cor}.
    \label{eq:shearingBoxEquations}
\end{equation}
These equations are equivalent to the standard non-ideal MHD equations with the exception of the two new source terms $\bm S_{\rm grav}$ and $\bm S_{\rm cor}$ that arise due to the non-inertial coordinate system. They describe the influence of an effective tidal force and of the Coriolis force:
\begin{equation}
    \bm S_{\rm grav}  = \begin{pmatrix}
   0 \\
   \rho \Omega_0^2 \left(2q x \bm\hat{e}_x\right)  \\
    \rho \Omega_0^2 \bm v \cdot \left(2q x \bm\hat{e}_x\right)   \\
   0\\
   \end{pmatrix}, \;\;\;\;\;\;  
   \bm S_{\rm cor} =  \begin{pmatrix}
   0\\
   -2 \rho \Omega_0  \bm\hat{e}_z \times \bm v\\
  0\\
  0
   \end{pmatrix}.
   \label{eq:shearingBoxSourceTerms}
\end{equation}
$\bm S_{\rm grav}$ depends on the shearing parameter
\begin{equation}
    q = - \frac{d \ln \Omega}{d \ln r},
\end{equation}
which simplifies to $q=3/2$ for the Keplerian case that we exclusively discuss in the present paper. We also neglect in this study the stratification in the $z$-direction of the disk, which would otherwise require an additional term in $\bm S_{\rm grav}$ in the vertical direction.

The above system allows for a ground-state solution with the velocity field
\begin{equation}
    \bm v_0 = (0,-q\Omega_0 x,0),
    \label{eq:backroundShearFlow}
\end{equation}
at constant pressure, constant density, and without a magnetic field. To close the system of equations we also have to define boundary conditions (BCs).  In the $y$-direction, we use standard periodic BCs, and in the $z$-direction periodic BCs as well. In the $x$-direction, we use the so-called shearing box BCs that are similar to standard periodic BCs but take into account the background shear flow as follows (\ref{eq:backroundShearFlow}):
\begin{subequations}
\begin{equation}
    f(x,y,z,t) = f(x \pm L_x, y \mp w t, z,t),\;\;\;\; f\in \{\rho, \rho v_x, \rho v_z, \bm B\},
\end{equation}
\begin{equation}
    \rho v_y(x,y,z,t) = \rho v_y(x \pm L_x, y \mp w t, z,t) \mp \rho  w ,
\end{equation}
\begin{equation}
    e(x,y,z,t) = e(x \pm L_x, y \mp w t, z,t) \mp \rho v_y v_w + \frac{\rho w^2}{2} ,
\end{equation}%
\label{eq:shearingBoxBoundaryConditions}%
\end{subequations}
where $L_x$ is the box size in the $x$-direction and $w = q \Omega_0 L_x$. The boundary conditions, therefore, do not conserve the azimuthal momentum, nor the total energy or the azimuthal component of the volume-weighted averaged magnetic field \citep{gressel2007shearingbox}:
\begin{equation}
  \frac{\partial \left < B_y\right>}{\partial t}  = -\frac{w}{V}\int_{\partial x} B_x\, {\rm d}y \,{\rm d}z.
  \label{eq:evolutionBy}
\end{equation}
Here $\partial x$ denotes the boundary in the $x$-direction, and $V$ is the total volume of the box. Only in the case that the magnetic field has no mean radial component and $\nabla \cdot \bm B = 0$ holds, the azimuthal field is conserved. 
We refer to \cite{zier2022simulating} for the details of the implementation of the shearing box in {\small AREPO}.

\subsection{The MRI dispersion relation with dissipation}
\label{subsec:nonidealMHDDispersion}

We assume that there is a constant mean field in the vertical and azimuthal direction, given by $\bm B = B_{0,y} \hat{e}_y + B_{0,z} \hat{e}_z$, and an isothermal equation of state. We then introduce small perturbations of the form $\bm b e^{ i \bm k \cdot \bm x + \sigma t}$, $\bm v e^{ i \bm k \cdot \bm x + \sigma t}$, and $\rho e^{i \bm k \cdot \bm x + \sigma t}$ to the magnetic field, velocity, and density $\rho_0$, respectively. Here, $\bm k = k_x \hat{e}_x + k_z \hat{e}_z$ represents the wavevector, and $\sigma$ denotes the growth rate of these perturbations.

We can linearize equation (\ref{eq:shearingBoxEquations}) around the ground state and substitute our Ansatz for the perturbations. This yields a system of linear equations, which we can solve to determine the eigenvectors and corresponding eigenvalues. The eigenvalues depend on the Alfvén frequency $\omega_A = \bm k \cdot \bm v_{A}$, the Alfvén velocity $\bm v_A = \bm B /\sqrt{\rho}$, and the Ohmic and ambipolar Elsasser numbers $\Lambda_{\rm OR/AD}$ (see equation \ref{eq:elssaserNumber}). This linear dispersion analysis and a brief accompanying discussion can be found in Appendix~\ref{app:fullDispersionRelatio}. Additionally, we provide in Appendix~\ref{app:fullDispersionRelatio} the numerical data of the eigenvalues and eigenvectors for the simulations discussed in the following.

\subsection{Ohmic diffusion regime}

As in the ideal MHD case, the fastest growing modes of the MRI share the property $k_x = 0$ if we only consider Ohmic diffusion and a purely vertical magnetic field ($B_{0,y} = 0$). As one can see in \cref{fig:linear_growth_ohmic}, the growth rate of the MRI monotonically decreases if one decreases the Elsasser number $\Lambda_{\rm OR}$. The suppression of the MRI occurs first for short wavelengths, resulting in a decrease in the wave number of the fastest-growing modes if one reduces $\Lambda_{\rm OR}$. To validate our simulation code, we have set up a simulation box of size $L_x \times L_y \times L_z = 1\times 1 \times 1$ and calculate the maximum growth rate $\sigma = 0.75 \Omega_0$ and the corresponding $\omega_A = 0.9682  \Omega_0$ for the case of ideal MHD. We fix this $\omega_A$ and change the Elsasser number to analyse the damping of the MRI by Ohmic diffusion. The size of the vertical background field was chosen to ensure that the fastest-growing mode had a wavelength of $\lambda=1$, fitting perfectly within our simulation box. This coresponds to a plasma $\beta = \rho_0 c_s^2 / B^2 = 84.2199$, and we choose as natural units $\Omega_0 = 1$ and $c_s =1$. For the initial conditions, we calculate the eigenvector which corresponds to the fastest growing mode, and normalize it so that  $|\bm b| /B_0 = 10^{-3}$ holds. The simulations were run until the magnetic perturbation had grown by a factor of around $O(20)$.  At the corresponding time $t_e$ we measured the average growth rate of the total magnetic field, denoted as $\sigma_{\rm sim}$, and we defined the error in the growth rate as $e = \left|\sigma_{\rm theory} - \sigma_{\rm sim}\right| /\sigma_{\rm theory}$.  

To examine the effect of resolution on our results, we conducted simulations at varying resolutions and plot in \cref{fig:linear_growth_ohmic_results} the error $e$ as a function of resolution for different Elsasser numbers. In all cases, the relative error becomes less than 1\% for high resolution, which means both codes can accurately describe the exponential growth of the MRI. For ideal MHD as well as only weak diffusion, the convergence of both codes is better than second order but approaches first order in the diffusion dominated regime ($\Lambda_{\rm OR} = 0.8$).

\subsection{Ambipolar diffusion regime}

For a purely vertical magnetic field, ambipolar diffusion works in the same way as Ohmic diffusion, so in this section we will add an azimuthal magnetic field $B_{0,y} = B_{0,z}$. In this case so-called oblique modes ($k_x \neq 0$) can grow fastest \citep{kunz2004ambipolar}. This can also be seen in \cref{fig:linear_growth_kx_ambipolar}, where we plot the growth rate as a function of $\omega_A = \bm k \cdot \bm B$ and $k_x / k_z$.  Interestingly, even for large $\omega_A$ (strong field or small wavelength) the oblique modes can be unstable, in contrast to MRI with ideal MHD. For simplicity, we choose $k_x = -k_z = 2 \pi$ for our tests and, as in the last section, we use a box of size $L_x \times L_y \times L_z = 1\times 1 \times 1$. For the background magnetic field we choose $\beta = 84.918$, which ensures that for ideal MHD our chosen mode is the fastest growing one with $k_x = -k_z$. We also use natural units $\Omega_0 = 1$ and $c_s = 1$, and again measure the error $e$ of the MRI growth rate as a function of resolution and ambipolar Elsasser number $\Lambda_{\rm AD}$. 

The results of these tests are shown in \cref{fig:linear_growth_ambipolar_results}. Both codes are able to resolve the growth of the MRI accurately for high resolutions, but in general the error increases when we decrease $\Lambda_{\rm AD}$. {\small AREPO} shows approximately third order convergence in $e$ except for the smallest $\Lambda_{\rm AD}$ at the highest resolution. For lower resolutions, {\small ATHENA++} shows smaller errors, but even for the ideal MHD it only converges to about first order. This could be explained by the oblique modes, which lead to fluxes that are not perpendicular to the Cartesian grid and can produce grid misalignment effects.  These are smaller in {\small AREPO}, which could explain its better convergence in this problem.

\begin{figure*}
    \centering
    \includegraphics[width=1\linewidth]{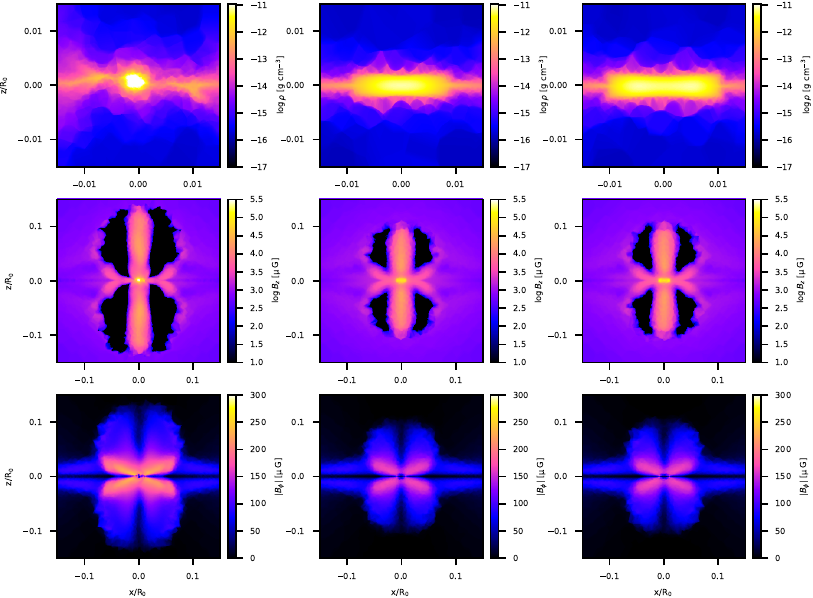}
    \caption{Formation of a dense core in the collapse of a magnetized cloud under its own self-gravity with different models for the magnetic field. The left column shows a simulation with ideal MHD while the middle one adds Ohmic diffusion and the right one ambipolar diffusion. All the panels show projected slices in the $xz$-plane with the slice having a thickness of 0.1 times the side length of the plot. In the top row, we display a volume-weighted density projection of the gas in the central region (0.06 $R_0$ in diameter). The middle and bottom row show density-weighted projections of the $z$- and the azimuthal component of the magnetic field, respectively, in a region 10 times as large. Note that the $z$-component of the magnetic field is negative (inverted with respect to the initial condition) due to the outflow in some regions. Results from the three simulations are shown from snapshots after an identical amount of elapsed time, approximately $3.4 \times 10^4$  years.
    }
    \label{fig:Collapse}
\end{figure*}

\section{Magnetized cloud collapse}
\label{sec:magnetizedCloudCollapse}

To test our new module for non-ideal MHD in an environment closer to  typical science applications, we study in this section the collapse of a rotating, magnetized cloud under its own gravity by performing the same simulation three times, with ideal MHD, with Ohmic diffusion, and with ambipolar diffusion. The chosen setup is important since it can act as a simplified model for star formation, allowing us to test the stability of our methods also in situations where a large range of densities and thus cell sizes occurs.

We use the same initial conditions as \cite{pakmor2011magnetohydrodynamics} and \cite{marinacci2018non}, which were in turn adapted from \cite{hennebelle2008magnetic}. They consist of a rigidly rotating, homogeneous sphere with radius $R_0 = 0.015\,\mathrm{pc}$ and mass $1\,{\rm M}_\odot$ embedded via a small transition region in a low density environment with a density contrast of $100$. The total size of the simulation box is $0.06\,\mathrm{pc}$. The initial density of the cloud is $4.8 \times 10^{-18} \mathrm{g\ cm^{-3}}$, which translates to a free fall time of $3 \times 10^4$ years. We choose a ratio of $0.045$ between rotational and gravitational energy, which is equivalent to a rotational period of $4.7 \times 10^5$ years for the rigid body rotation around the $z-$axis. The cloud is initially penetrated by a homogeneous, purely vertical magnetic field of magnitude $107\,\mathrm{\mu G}$, equivalent to a mass-to-flux over critical mass-to-flux ratio of $5.6$. We use a barotropic equation of state \citep{hennebelle2008magnetic}
\begin{equation}
    P = \rho c_{s,0}^2 \sqrt{1 + \left(\rho / \rho_c \right)^{4/3}},
\end{equation}
with sound speed $c_{s,0} = 0.2\, \mathrm{km\ s^{-1}}$ and critical density $\rho_c = 10^{-13}\,\mathrm{g\ cm^{-3}}$, which represents a transition from an initially isothermal EOS to an adiabatic one at higher densities. We start with an initial resolution of $128^3$ cells on a uniform Cartesian mesh and apply periodic boundary conditions at the edges of the simulation box for all quantities except gravity. Cells are refined if their free fall time is smaller than ten times their sound crossing time. We also introduce a minimum volume of $5 \times 10^{-17} \mathrm{pc}$ which is equivalent to an effective resolution of $16384^3$ cells.  The latter condition reduces the computational cost, especially considering the quadratic dependence of the diffusive timestep on cell radius. 

In the simulations including one of the diffusive effects, we set a constant diffusivity of $\eta_{\mathrm{OR,AD}} = 10^{18} \mathrm{cm^2\ s^{-1}}$. As noted above, these coefficients actually vary with density and magnetic field, and ambipolar diffusion is generally much stronger than Ohmic diffusion, except in the innermost core \citep[e.g.][]{wurster2016nicil}. However, our choice allows for a closer comparison with the results from other numerical implementations.  

After about $3.4 \times 10^4$ years of simulated evolution we find around $2.2$ million cells, about $\approx 10^5$ cells more than in the initial conditions. Most of the cells are in and around the dense core that formed, and which we show in \cref{fig:Collapse}. While this core is approximately spherical in the case of ideal MHD, the simulations including one of the diffusive effects form a more extended disc structure instead. Differences can also be seen in the magnetic field, which reaches higher strengths (more than $\sim 10^{5}\,\mathrm{\mu G}$) in the innermost region in the ideal case and traces outflows from the central region that are launched with a higher velocity than in the diffusive runs, having already reached a larger distance from the core. The two simulations with non-ideal effects show very similar behavior, which is expected, as the additional terms in the induction equation~(\ref{eq:nonidelMHDSourceTerms}) give identical results if the electrical current and the magnetic field are orthogonal, which is approximately the case throughout most of the simulation. However, there are differences in the small-scale azimuthal field, which builds up due to the rotation of the gas and gives a magnetic field component which is parallel to the mostly azimuthal current. As the drift of field lines in ambipolar diffusion compared to ideal MHD  is strongest when the current and the magnetic field are perpendicular, ambipolar diffusion is slightly less effective in reducing the azimuthal field than Ohmic diffusion. 

Our results for the collapse with ideal MHD and Ohmic diffusion are very similar to  previous ones by \cite{marinacci2018non}\footnote{This paper contained a typo and used an initial magnetic field strength of $107\,\mathrm{\mu G}$, not $30\,\mathrm{\mu G}$ as stated. However, the fields displayed in the plots were falsely multiplied by the corresponding factor. There was also a sign error, which means they showed in their Fig. 11 the radial component of the magnetic field instead of the azimuthal one.}, showing that our new implementation is similar to the earlier method in the case of Ohmic diffusion. The formation of rotation-supported  disks in our non-ideal MHD simulations is an example of how diffusion of the magnetic field can solve the magnetic braking catastrophe, as seen before, e.g., by  \cite{dapp_basu2010} and \cite{vaytet2018non_ideal}. The reduction of the speed of the outflow in non-ideal MHD agrees with earlier work of \cite{wurster2018outflows}. Lower maximum values of the magnetic field are a consistent consequence of the inclusion of non-ideal MHD, as observed for example by \cite{masson2016} and \cite{wurster2021}. While all the cited works include non-constant values for the diffusion coefficients, we have shown that we can qualitatively reproduce these trends in our simplified model.

\section{Summary and Conclusions}
\label{sec:discussionSummary}

In this work, we have extended the magnetohydrodynamic moving mesh code {\small AREPO} by incorporating a new diffusion solver for Ohmic and ambipolar diffusion. Both effects become important in cold and dense environments, in which the gas does not perfectly couple to the magnetic field. Prominent examples are protostellar cores or protoplanetary disks, which can also profit from the adaptive resolution and Lagrangian nature of the moving mesh method.

The core of our method is an accurate gradient estimate by performing an overdetermined least square fit (LSF) at the centre of interfaces between neighbouring Voronoi cells. This algorithm is more stable than the corner-based method introduced in \cite{pakmor2016semi} and was further improved by, e.g., adding a simple preconditioner for the LSF. We also added a further limiter that uses a finite difference estimate as a fallback if the LSF becomes a strong extrapolation, which reduces the risk of introducing new extrema.

We have tested our implementation for several setups with known analytical or semi-analytical solutions: The pure diffusion of a magnetic field, the structure of a magnetic shock with diffusion (C-shock) and the damping of an Alvfen wave. In all cases, the accuracy for a given resolution was comparable to results obtained with the state-of-the-art static grid code {\small ATHENA++} with the exception of the adiabatic C-shock test. In this case, we observed noise in the post-shock region but this is an artifact which can also be observed in the ideal MHD case and does not originate from the nonideal MHD solver itself. We have also shown that our new method can accurately describe the linear growth of the magnetorotational instability (MRI) in a shearing box, even in the diffusion-dominated regime. As a first example of a more complicated setup, we have simulated the collapse of a magnetized gas cloud with constant Ohmic and ambipolar diffusion. Consistent with earlier simulations of this problem, we find that the magnetic diffusion terms reduce the magnetic field amplification and support the creation of a disk. This test also demonstrates that our new scheme is robust for large density contrasts.

We are currently working on applying the code to more realistic simulations of the collapse of magnetized protostellar cores with non-constant diffusivities (Mayer et al., in preparation). This is a  promising application area for our new method since the quasi-Lagrangian nature of {\small AREPO} can automatically increase the spatial resolution in the forming core. 

We also expect our approach to be particularly useful for non-axisymmetric setups where the rotation and magnetic field axis are misaligned, or when there is turbulence already in the initial conditions. We note that in star-formation simulations also the Hall effect, the third non-ideal MHD effect, plays a role. It is not diffusive but dispersive in character, and introduces the whistler waves as a new MHD wave, making numerical  implementations more difficult. In fact, it is well known that numerical treatments of the Hall effect can easily become unstable \citep{kunz2013}.  We will discuss in a forthcoming work how we can incorporate it in {\small AREPO} (Zier et al., , in preparation) as an extension to our current methods.

\section*{Acknowledgements}

The authors acknowledge helpful discussions with R\"udiger Pakmor.
We would also like to thank Federico Marinacci and Mark Vogelsberger for giving us access to their simulation data for the magnetized cloud collapse.

\section*{Data Availability}
The data underlying this paper will be shared upon reasonable request to the corresponding author.

\begin{appendix}

\renewcommand{\thefigure}{A\arabic{figure}}
\setcounter{figure}{0}

\bibliographystyle{mnras}
\bibliography{main.bib}

\begin{thebibliography}{}
\makeatletter
\relax
\def\mn@urlcharsother{\let\do\@makeother \do\$\do\&\do\#\do\^\do\_\do\%\do\~}
\def\mn@doi{\begingroup\mn@urlcharsother \@ifnextchar [ {\mn@doi@}
  {\mn@doi@[]}}
\def\mn@doi@[#1]#2{\def\@tempa{#1}\ifx\@tempa\@empty \href
  {http://dx.doi.org/#2} {doi:#2}\else \href {http://dx.doi.org/#2} {#1}\fi
  \endgroup}
\def\mn@eprint#1#2{\mn@eprint@#1:#2::\@nil}
\def\mn@eprint@arXiv#1{\href {http://arxiv.org/abs/#1} {{\tt arXiv:#1}}}
\def\mn@eprint@dblp#1{\href {http://dblp.uni-trier.de/rec/bibtex/#1.xml}
  {dblp:#1}}
\def\mn@eprint@#1:#2:#3:#4\@nil{\def\@tempa {#1}\def\@tempb {#2}\def\@tempc
  {#3}\ifx \@tempc \@empty \let \@tempc \@tempb \let \@tempb \@tempa \fi \ifx
  \@tempb \@empty \def\@tempb {arXiv}\fi \@ifundefined
  {mn@eprint@\@tempb}{\@tempb:\@tempc}{\expandafter \expandafter \csname
  mn@eprint@\@tempb\endcsname \expandafter{\@tempc}}}

\bibitem[\protect\citeauthoryear{{Allen}, {Li}  \& {Shu}}{{Allen}
  et~al.}{2003}]{Allen2003}
{Allen} A.,  {Li} Z.-Y.,   {Shu} F.~H.,  2003, \mn@doi [\apj] {10.1086/379243},
  \href {https://ui.adsabs.harvard.edu/abs/2003ApJ...599..363A} {599, 363}

\bibitem[\protect\citeauthoryear{{Bai}}{{Bai}}{2014}]{bai2014hall}
{Bai} X.-N.,  2014, \mn@doi [\apj] {10.1088/0004-637X/791/2/137}, \href
  {https://ui.adsabs.harvard.edu/abs/2014ApJ...791..137B} {791, 137}

\bibitem[\protect\citeauthoryear{{Bai} \& {Stone}}{{Bai} \&
  {Stone}}{2011}]{Bai2011}
{Bai} X.-N.,  {Stone} J.~M.,  2011, \mn@doi [\apj]
  {10.1088/0004-637X/736/2/144}, \href
  {https://ui.adsabs.harvard.edu/abs/2011ApJ...736..144B} {736, 144}

\bibitem[\protect\citeauthoryear{{Balbus} \& {Hawley}}{{Balbus} \&
  {Hawley}}{1991}]{balbus1991powerful}
{Balbus} S.~A.,  {Hawley} J.~F.,  1991, \mn@doi [\apj] {10.1086/170270}, \href
  {https://ui.adsabs.harvard.edu/abs/1991ApJ...376..214B} {376, 214}

\bibitem[\protect\citeauthoryear{{Basu} \& {Ciolek}}{{Basu} \&
  {Ciolek}}{2004}]{basu2004formation}
{Basu} S.,  {Ciolek} G.~E.,  2004, \mn@doi [\apjl] {10.1086/421464}, \href
  {https://ui.adsabs.harvard.edu/abs/2004ApJ...607L..39B} {607, L39}

\bibitem[\protect\citeauthoryear{{Basu} \& {Dapp}}{{Basu} \&
  {Dapp}}{2010}]{basu2010long}
{Basu} S.,  {Dapp} W.~B.,  2010, \mn@doi [\apj] {10.1088/0004-637X/716/1/427},
  \href {https://ui.adsabs.harvard.edu/abs/2010ApJ...716..427B} {716, 427}

\bibitem[\protect\citeauthoryear{{Bate}}{{Bate}}{2011}]{Bate2011}
{Bate} M.~R.,  2011, \mn@doi [\mnras] {10.1111/j.1365-2966.2011.19386.x}, \href
  {https://ui.adsabs.harvard.edu/abs/2011MNRAS.417.2036B} {417, 2036}

\bibitem[\protect\citeauthoryear{{Bate}, {Tricco}  \& {Price}}{{Bate}
  et~al.}{2014}]{bate2014collapse}
{Bate} M.~R.,  {Tricco} T.~S.,   {Price} D.~J.,  2014, \mn@doi [\mnras]
  {10.1093/mnras/stt1865}, \href
  {https://ui.adsabs.harvard.edu/abs/2014MNRAS.437...77B} {437, 77}

\bibitem[\protect\citeauthoryear{{Berlok}, {Pakmor}  \& {Pfrommer}}{{Berlok}
  et~al.}{2020}]{berlok2020braginskii}
{Berlok} T.,  {Pakmor} R.,   {Pfrommer} C.,  2020, \mn@doi [\mnras]
  {10.1093/mnras/stz3115}, \href
  {https://ui.adsabs.harvard.edu/abs/2020MNRAS.491.2919B} {491, 2919}

\bibitem[\protect\citeauthoryear{{Chandrasekhar}}{{Chandrasekhar}}{1960}]{chandrasekhar1960stability}
{Chandrasekhar} S.,  1960, \mn@doi [Proceedings of the National Academy of
  Science] {10.1073/pnas.46.2.253}, \href
  {https://ui.adsabs.harvard.edu/abs/1960PNAS...46..253C} {46, 253}

\bibitem[\protect\citeauthoryear{{Cheng}, {Guo}  \& {Ding}}{{Cheng}
  et~al.}{2017}]{Cheng2017}
{Cheng} X.,  {Guo} Y.,   {Ding} M.,  2017, \mn@doi [Science China Earth
  Sciences] {10.1007/s11430-017-9074-6}, \href
  {https://ui.adsabs.harvard.edu/abs/2017ScChD..60.1383C} {60, 1383}

\bibitem[\protect\citeauthoryear{{Choi}, {Kim}  \& {Wiita}}{{Choi}
  et~al.}{2009}]{choi2009explicit}
{Choi} E.,  {Kim} J.,   {Wiita} P.~J.,  2009, \mn@doi [\apjs]
  {10.1088/0067-0049/181/2/413}, \href
  {https://ui.adsabs.harvard.edu/abs/2009ApJS..181..413C} {181, 413}

\bibitem[\protect\citeauthoryear{{Cox}}{{Cox}}{2005}]{cox2005three}
{Cox} D.~P.,  2005, \mn@doi [\araa] {10.1146/annurev.astro.43.072103.150615},
  \href {https://ui.adsabs.harvard.edu/abs/2005ARA&A..43..337C} {43, 337}

\bibitem[\protect\citeauthoryear{{Dapp} \& {Basu}}{{Dapp} \&
  {Basu}}{2010}]{dapp_basu2010}
{Dapp} W.~B.,  {Basu} S.,  2010, \mn@doi [\aap] {10.1051/0004-6361/201015700},
  \href {https://ui.adsabs.harvard.edu/abs/2010A&A...521L..56D} {521, L56}

\bibitem[\protect\citeauthoryear{{Dedner}, {Kemm}, {Kr{\"o}ner}, {Munz},
  {Schnitzer}  \& {Wesenberg}}{{Dedner} et~al.}{2002}]{dedner2002hyperbolic}
{Dedner} A.,  {Kemm} F.,  {Kr{\"o}ner} D.,  {Munz} C.~D.,  {Schnitzer} T.,
  {Wesenberg} M.,  2002, \mn@doi [Journal of Computational Physics]
  {10.1006/jcph.2001.6961}, \href
  {https://ui.adsabs.harvard.edu/abs/2002JCoPh.175..645D} {175, 645}

\bibitem[\protect\citeauthoryear{{Dolag}, {Bartelmann}  \& {Lesch}}{{Dolag}
  et~al.}{1999}]{dolag1999sph}
{Dolag} K.,  {Bartelmann} M.,   {Lesch} H.,  1999, \mn@doi [\aap]
  {10.48550/arXiv.astro-ph/0202272}, \href
  {https://ui.adsabs.harvard.edu/abs/1999A&A...348..351D} {348, 351}

\bibitem[\protect\citeauthoryear{{Dolag}, {Bartelmann}  \& {Lesch}}{{Dolag}
  et~al.}{2002}]{dolag2002evolution}
{Dolag} K.,  {Bartelmann} M.,   {Lesch} H.,  2002, \mn@doi [\aap]
  {10.1051/0004-6361:20020241}, \href
  {https://ui.adsabs.harvard.edu/abs/2002A&A...387..383D} {387, 383}

\bibitem[\protect\citeauthoryear{{Dolag}, {Komatsu}  \& {Sunyaev}}{{Dolag}
  et~al.}{2016}]{dolag2016sz}
{Dolag} K.,  {Komatsu} E.,   {Sunyaev} R.,  2016, \mn@doi [\mnras]
  {10.1093/mnras/stw2035}, \href
  {https://ui.adsabs.harvard.edu/abs/2016MNRAS.463.1797D} {463, 1797}

\bibitem[\protect\citeauthoryear{{Draine}}{{Draine}}{1980}]{Draine1980}
{Draine} B.~T.,  1980, \mn@doi [\apj] {10.1086/158416}, \href
  {https://ui.adsabs.harvard.edu/abs/1980ApJ...241.1021D} {241, 1021}

\bibitem[\protect\citeauthoryear{{Duffin} \& {Pudritz}}{{Duffin} \&
  {Pudritz}}{2008}]{Duffin2008}
{Duffin} D.~F.,  {Pudritz} R.~E.,  2008, \mn@doi [\mnras]
  {10.1111/j.1365-2966.2008.14026.x}, \href
  {https://ui.adsabs.harvard.edu/abs/2008MNRAS.391.1659D} {391, 1659}

\bibitem[\protect\citeauthoryear{Elmegreen}{Elmegreen}{1979}]{elmegreen1979magnetic}
Elmegreen B.~G.,  1979, The Astrophysical Journal, 232, 729

\bibitem[\protect\citeauthoryear{{Evans} \& {Hawley}}{{Evans} \&
  {Hawley}}{1988}]{evans1988simulation}
{Evans} C.~R.,  {Hawley} J.~F.,  1988, \mn@doi [\apj] {10.1086/166684}, \href
  {https://ui.adsabs.harvard.edu/abs/1988ApJ...332..659E} {332, 659}

\bibitem[\protect\citeauthoryear{{Fermi}}{{Fermi}}{1949}]{fermi1949origin}
{Fermi} E.,  1949, \mn@doi [Physical Review] {10.1103/PhysRev.75.1169}, \href
  {https://ui.adsabs.harvard.edu/abs/1949PhRv...75.1169F} {75, 1169}

\bibitem[\protect\citeauthoryear{Ferriere}{Ferriere}{2001}]{ferriere2001interstellar}
Ferriere K.~M.,  2001, Reviews of Modern Physics, 73, 1031

\bibitem[\protect\citeauthoryear{{Fricke}}{{Fricke}}{1969}]{fricke1969stability}
{Fricke} K.,  1969, \aap, \href
  {https://ui.adsabs.harvard.edu/abs/1969A&A.....1..388F} {1, 388}

\bibitem[\protect\citeauthoryear{{Goldreich} \& {Lynden-Bell}}{{Goldreich} \&
  {Lynden-Bell}}{1965}]{goldreich1965ii}
{Goldreich} P.,  {Lynden-Bell} D.,  1965, \mn@doi [\mnras]
  {10.1093/mnras/130.2.125}, \href
  {https://ui.adsabs.harvard.edu/abs/1965MNRAS.130..125G} {130, 125}

\bibitem[\protect\citeauthoryear{{Gonz{\'a}lez-Morales}, {Khomenko}, {Downes}
  \& {de Vicente}}{{Gonz{\'a}lez-Morales} et~al.}{2018}]{gonzalez2018mhdsts}
{Gonz{\'a}lez-Morales} P.~A.,  {Khomenko} E.,  {Downes} T.~P.,   {de Vicente}
  A.,  2018, \mn@doi [\aap] {10.1051/0004-6361/201731916}, \href
  {https://ui.adsabs.harvard.edu/abs/2018A&A...615A..67G} {615, A67}

\bibitem[\protect\citeauthoryear{{Gressel} \& {Ziegler}}{{Gressel} \&
  {Ziegler}}{2007}]{gressel2007shearingbox}
{Gressel} O.,  {Ziegler} U.,  2007, \mn@doi [Comput. Phys. Commun.]
  {10.1016/j.cpc.2007.01.010}, \href
  {https://ui.adsabs.harvard.edu/abs/2007CoPhC.176..652G} {176, 652}

\bibitem[\protect\citeauthoryear{{Grudi{\'c}}, {Guszejnov}, {Hopkins}, {Offner}
   \& {Faucher-Gigu{\`e}re}}{{Grudi{\'c}} et~al.}{2021}]{grudic2021starforge}
{Grudi{\'c}} M.~Y.,  {Guszejnov} D.,  {Hopkins} P.~F.,  {Offner} S. S.~R.,
  {Faucher-Gigu{\`e}re} C.-A.,  2021, \mn@doi [\mnras]
  {10.1093/mnras/stab1347}, \href
  {https://ui.adsabs.harvard.edu/abs/2021MNRAS.506.2199G} {506, 2199}

\bibitem[\protect\citeauthoryear{Grundy \& McLaughlin}{Grundy \&
  McLaughlin}{1982}]{grundy1982eigenvalues}
Grundy R.,  McLaughlin R.,  1982, Proceedings of the Royal Society of London.
  A. Mathematical and Physical Sciences, 383, 89

\bibitem[\protect\citeauthoryear{{Hennebelle} \& {Teyssier}}{{Hennebelle} \&
  {Teyssier}}{2008}]{hennebelle2008magnetic}
{Hennebelle} P.,  {Teyssier} R.,  2008, \mn@doi [\aap]
  {10.1051/0004-6361:20078310}, \href
  {https://ui.adsabs.harvard.edu/abs/2008A&A...477...25H} {477, 25}

\bibitem[\protect\citeauthoryear{Hill}{Hill}{1878}]{hill1878collected}
Hill G.,  1878, J. Math, 1, 5

\bibitem[\protect\citeauthoryear{{Hopkins}}{{Hopkins}}{2017}]{hopkins2017anisotropic}
{Hopkins} P.~F.,  2017, \mn@doi [\mnras] {10.1093/mnras/stw3306}, \href
  {https://ui.adsabs.harvard.edu/abs/2017MNRAS.466.3387H} {466, 3387}

\bibitem[\protect\citeauthoryear{{Inoue} \& {Inutsuka}}{{Inoue} \&
  {Inutsuka}}{2008}]{inoue2008two}
{Inoue} T.,  {Inutsuka} S.-i.,  2008, \mn@doi [\apj] {10.1086/590528}, \href
  {https://ui.adsabs.harvard.edu/abs/2008ApJ...687..303I} {687, 303}

\bibitem[\protect\citeauthoryear{{Inoue} \& {Inutsuka}}{{Inoue} \&
  {Inutsuka}}{2009}]{inoue2009two}
{Inoue} T.,  {Inutsuka} S.-i.,  2009, \mn@doi [\apj]
  {10.1088/0004-637X/704/1/161}, \href
  {https://ui.adsabs.harvard.edu/abs/2009ApJ...704..161I} {704, 161}

\bibitem[\protect\citeauthoryear{{Kim} \& {Ostriker}}{{Kim} \&
  {Ostriker}}{2018}]{kim2018numerical}
{Kim} C.-G.,  {Ostriker} E.~C.,  2018, \mn@doi [\apj]
  {10.3847/1538-4357/aaa5ff}, \href
  {https://ui.adsabs.harvard.edu/abs/2018ApJ...853..173K} {853, 173}

\bibitem[\protect\citeauthoryear{{Kotera} \& {Olinto}}{{Kotera} \&
  {Olinto}}{2011}]{kotera2011astrophysics}
{Kotera} K.,  {Olinto} A.~V.,  2011, \mn@doi [\araa]
  {10.1146/annurev-astro-081710-102620}, \href
  {https://ui.adsabs.harvard.edu/abs/2011ARA&A..49..119K} {49, 119}

\bibitem[\protect\citeauthoryear{{Kunz} \& {Balbus}}{{Kunz} \&
  {Balbus}}{2004}]{kunz2004ambipolar}
{Kunz} M.~W.,  {Balbus} S.~A.,  2004, \mn@doi [\mnras]
  {10.1111/j.1365-2966.2004.07383.x}, \href
  {https://ui.adsabs.harvard.edu/abs/2004MNRAS.348..355K} {348, 355}

\bibitem[\protect\citeauthoryear{{Kunz} \& {Lesur}}{{Kunz} \&
  {Lesur}}{2013}]{kunz2013}
{Kunz} M.~W.,  {Lesur} G.,  2013, \mn@doi [\mnras] {10.1093/mnras/stt1171},
  \href {https://ui.adsabs.harvard.edu/abs/2013MNRAS.434.2295K} {434, 2295}

\bibitem[\protect\citeauthoryear{{Lesur}}{{Lesur}}{2021}]{lesur2021magnetohydrodynamics}
{Lesur} G.,  2021, \mn@doi [Journal of Plasma Physics]
  {10.1017/S0022377820001002}, \href
  {https://ui.adsabs.harvard.edu/abs/2021JPlPh..87a2001P} {87, 205870101}

\bibitem[\protect\citeauthoryear{{Lesur}, {Kunz}  \& {Fromang}}{{Lesur}
  et~al.}{2014}]{lesur2014thanatology}
{Lesur} G.,  {Kunz} M.~W.,   {Fromang} S.,  2014, \mn@doi [\aap]
  {10.1051/0004-6361/201423660}, \href
  {https://ui.adsabs.harvard.edu/abs/2014A&A...566A..56L} {566, A56}

\bibitem[\protect\citeauthoryear{{Li}, {McKee}, {Klein}  \& {Fisher}}{{Li}
  et~al.}{2008}]{li2008sub}
{Li} P.~S.,  {McKee} C.~F.,  {Klein} R.~I.,   {Fisher} R.~T.,  2008, \mn@doi
  [\apj] {10.1086/589874}, \href
  {https://ui.adsabs.harvard.edu/abs/2008ApJ...684..380L} {684, 380}

\bibitem[\protect\citeauthoryear{{Li}, {Krasnopolsky}  \& {Shang}}{{Li}
  et~al.}{2011}]{li2011non}
{Li} Z.-Y.,  {Krasnopolsky} R.,   {Shang} H.,  2011, \mn@doi [\apj]
  {10.1088/0004-637X/738/2/180}, \href
  {https://ui.adsabs.harvard.edu/abs/2011ApJ...738..180L} {738, 180}

\bibitem[\protect\citeauthoryear{Lynden-Bell \& Pringle}{Lynden-Bell \&
  Pringle}{1974}]{lynden1974evolution}
Lynden-Bell D.,  Pringle J.~E.,  1974, \mnras, 168, 603

\bibitem[\protect\citeauthoryear{{Mac Low}, {Norman}, {Konigl}  \&
  {Wardle}}{{Mac Low} et~al.}{1995}]{MacLow1995}
{Mac Low} M.-M.,  {Norman} M.~L.,  {Konigl} A.,   {Wardle} M.,  1995, \mn@doi
  [\apj] {10.1086/175477}, \href
  {https://ui.adsabs.harvard.edu/abs/1995ApJ...442..726M} {442, 726}

\bibitem[\protect\citeauthoryear{{Marchand}, {Commer{\c{c}}on}  \&
  {Chabrier}}{{Marchand} et~al.}{2018}]{marchand2018impact}
{Marchand} P.,  {Commer{\c{c}}on} B.,   {Chabrier} G.,  2018, \mn@doi [\aap]
  {10.1051/0004-6361/201832907}, \href
  {https://ui.adsabs.harvard.edu/abs/2018A&A...619A..37M} {619, A37}

\bibitem[\protect\citeauthoryear{{Marinacci} \& {Vogelsberger}}{{Marinacci} \&
  {Vogelsberger}}{2016}]{marinacci2015effects}
{Marinacci} F.,  {Vogelsberger} M.,  2016, \mn@doi [\mnras]
  {10.1093/mnrasl/slv176}, \href
  {https://ui.adsabs.harvard.edu/abs/2016MNRAS.456L..69M} {456, L69}

\bibitem[\protect\citeauthoryear{{Marinacci}, {Vogelsberger}, {Mocz}  \&
  {Pakmor}}{{Marinacci} et~al.}{2015}]{marinacci2015large}
{Marinacci} F.,  {Vogelsberger} M.,  {Mocz} P.,   {Pakmor} R.,  2015, \mn@doi
  [\mnras] {10.1093/mnras/stv1692}, \href
  {https://ui.adsabs.harvard.edu/abs/2015MNRAS.453.3999M} {453, 3999}

\bibitem[\protect\citeauthoryear{{Marinacci}, {Vogelsberger}, {Kannan}, {Mocz},
  {Pakmor}  \& {Springel}}{{Marinacci} et~al.}{2018a}]{marinacci2018non}
{Marinacci} F.,  {Vogelsberger} M.,  {Kannan} R.,  {Mocz} P.,  {Pakmor} R.,
  {Springel} V.,  2018a, \mn@doi [\mnras] {10.1093/mnras/sty397}, \href
  {https://ui.adsabs.harvard.edu/abs/2018MNRAS.476.2476M} {476, 2476}

\bibitem[\protect\citeauthoryear{{Marinacci} et~al.,}{{Marinacci}
  et~al.}{2018b}]{marinacci2018first}
{Marinacci} F.,  et~al., 2018b, \mn@doi [\mnras] {10.1093/mnras/sty2206}, \href
  {https://ui.adsabs.harvard.edu/abs/2018MNRAS.480.5113M} {480, 5113}

\bibitem[\protect\citeauthoryear{{Masson}, {Teyssier}, {Mulet-Marquis},
  {Hennebelle}  \& {Chabrier}}{{Masson} et~al.}{2012}]{masson2012incorporating}
{Masson} J.,  {Teyssier} R.,  {Mulet-Marquis} C.,  {Hennebelle} P.,
  {Chabrier} G.,  2012, \mn@doi [\apjs] {10.1088/0067-0049/201/2/24}, \href
  {https://ui.adsabs.harvard.edu/abs/2012ApJS..201...24M} {201, 24}

\bibitem[\protect\citeauthoryear{{Masson}, {Chabrier}, {Hennebelle}, {Vaytet}
  \& {Commer{\c{c}}on}}{{Masson} et~al.}{2016}]{masson2016}
{Masson} J.,  {Chabrier} G.,  {Hennebelle} P.,  {Vaytet} N.,
  {Commer{\c{c}}on} B.,  2016, \mn@doi [\aap] {10.1051/0004-6361/201526371},
  \href {https://ui.adsabs.harvard.edu/abs/2016A&A...587A..32M} {587, A32}

\bibitem[\protect\citeauthoryear{{Mellon} \& {Li}}{{Mellon} \&
  {Li}}{2008}]{mellon2008magnetic}
{Mellon} R.~R.,  {Li} Z.-Y.,  2008, \mn@doi [\apj] {10.1086/587542}, \href
  {https://ui.adsabs.harvard.edu/abs/2008ApJ...681.1356M} {681, 1356}

\bibitem[\protect\citeauthoryear{{Mocz}, {Vogelsberger}  \& {Hernquist}}{{Mocz}
  et~al.}{2014}]{mocz2014constrained}
{Mocz} P.,  {Vogelsberger} M.,   {Hernquist} L.,  2014, \mn@doi [\mnras]
  {10.1093/mnras/stu865}, \href
  {https://ui.adsabs.harvard.edu/abs/2014MNRAS.442...43M} {442, 43}

\bibitem[\protect\citeauthoryear{{Ntormousi}, {Hennebelle}, {Andr{\'e}}  \&
  {Masson}}{{Ntormousi} et~al.}{2016}]{ntormousi2016effect}
{Ntormousi} E.,  {Hennebelle} P.,  {Andr{\'e}} P.,   {Masson} J.,  2016,
  \mn@doi [\aap] {10.1051/0004-6361/201527400}, \href
  {https://ui.adsabs.harvard.edu/abs/2016A&A...589A..24N} {589, A24}

\bibitem[\protect\citeauthoryear{{O'Sullivan} \& {Downes}}{{O'Sullivan} \&
  {Downes}}{2006}]{o2006explicit}
{O'Sullivan} S.,  {Downes} T.~P.,  2006, \mn@doi [\mnras]
  {10.1111/j.1365-2966.2005.09898.x}, \href
  {https://ui.adsabs.harvard.edu/abs/2006MNRAS.366.1329O} {366, 1329}

\bibitem[\protect\citeauthoryear{{Pakmor} \& {Springel}}{{Pakmor} \&
  {Springel}}{2013}]{pakmor2013simulations}
{Pakmor} R.,  {Springel} V.,  2013, \mn@doi [\mnras] {10.1093/mnras/stt428},
  \href {https://ui.adsabs.harvard.edu/abs/2013MNRAS.432..176P} {432, 176}

\bibitem[\protect\citeauthoryear{{Pakmor}, {Bauer}  \& {Springel}}{{Pakmor}
  et~al.}{2011}]{pakmor2011magnetohydrodynamics}
{Pakmor} R.,  {Bauer} A.,   {Springel} V.,  2011, \mn@doi [\mnras]
  {10.1111/j.1365-2966.2011.19591.x}, \href
  {https://ui.adsabs.harvard.edu/abs/2011MNRAS.418.1392P} {418, 1392}

\bibitem[\protect\citeauthoryear{{Pakmor}, {Marinacci}  \& {Springel}}{{Pakmor}
  et~al.}{2014}]{pakmor2014magnetic}
{Pakmor} R.,  {Marinacci} F.,   {Springel} V.,  2014, \mn@doi [\apjl]
  {10.1088/2041-8205/783/1/L20}, \href
  {https://ui.adsabs.harvard.edu/abs/2014ApJ...783L..20P} {783, L20}

\bibitem[\protect\citeauthoryear{{Pakmor}, {Springel}, {Bauer}, {Mocz},
  {Munoz}, {Ohlmann}, {Schaal}  \& {Zhu}}{{Pakmor}
  et~al.}{2016a}]{pakmor2016improving}
{Pakmor} R.,  {Springel} V.,  {Bauer} A.,  {Mocz} P.,  {Munoz} D.~J.,
  {Ohlmann} S.~T.,  {Schaal} K.,   {Zhu} C.,  2016a, \mn@doi [\mnras]
  {10.1093/mnras/stv2380}, \href
  {https://ui.adsabs.harvard.edu/abs/2016MNRAS.455.1134P} {455, 1134}

\bibitem[\protect\citeauthoryear{{Pakmor}, {Pfrommer}, {Simpson}, {Kannan}  \&
  {Springel}}{{Pakmor} et~al.}{2016b}]{pakmor2016semi}
{Pakmor} R.,  {Pfrommer} C.,  {Simpson} C.~M.,  {Kannan} R.,   {Springel} V.,
  2016b, \mn@doi [\mnras] {10.1093/mnras/stw1761}, \href
  {https://ui.adsabs.harvard.edu/abs/2016MNRAS.462.2603P} {462, 2603}

\bibitem[\protect\citeauthoryear{{Pakmor} et~al.,}{{Pakmor}
  et~al.}{2017}]{pakmor2017magnetic}
{Pakmor} R.,  et~al., 2017, \mn@doi [\mnras] {10.1093/mnras/stx1074}, \href
  {https://ui.adsabs.harvard.edu/abs/2017MNRAS.469.3185P} {469, 3185}

\bibitem[\protect\citeauthoryear{{Parker}}{{Parker}}{1983}]{parker1983magnetic}
{Parker} E.~N.,  1983, \mn@doi [\apj] {10.1086/160637}, \href
  {https://ui.adsabs.harvard.edu/abs/1983ApJ...264..642P} {264, 642}

\bibitem[\protect\citeauthoryear{{Powell}, {Roe}, {Linde}, {Gombosi}  \& {De
  Zeeuw}}{{Powell} et~al.}{1999}]{powell1999solution}
{Powell} K.~G.,  {Roe} P.~L.,  {Linde} T.~J.,  {Gombosi} T.~I.,   {De Zeeuw}
  D.~L.,  1999, \mn@doi [Journal of Computational Physics]
  {10.1006/jcph.1999.6299}, \href
  {https://ui.adsabs.harvard.edu/abs/1999JCoPh.154..284P} {154, 284}

\bibitem[\protect\citeauthoryear{{Price} et~al.,}{{Price}
  et~al.}{2018}]{price2018phantom}
{Price} D.~J.,  et~al., 2018, \mn@doi [\pasa] {10.1017/pasa.2018.25}, \href
  {https://ui.adsabs.harvard.edu/abs/2018PASA...35...31P} {35, e031}

\bibitem[\protect\citeauthoryear{{Rodgers-Lee}, {Ray}  \&
  {Downes}}{{Rodgers-Lee} et~al.}{2016}]{rodgers2016global}
{Rodgers-Lee} D.,  {Ray} T.~P.,   {Downes} T.~P.,  2016, \mn@doi [\mnras]
  {10.1093/mnras/stw1980}, \href
  {https://ui.adsabs.harvard.edu/abs/2016MNRAS.463..134R} {463, 134}

\bibitem[\protect\citeauthoryear{{Shakura} \& {Sunyaev}}{{Shakura} \&
  {Sunyaev}}{1973}]{shakura1973black}
{Shakura} N.~I.,  {Sunyaev} R.~A.,  1973, \aap, \href
  {https://ui.adsabs.harvard.edu/abs/1973A&A....24..337S} {24, 337}

\bibitem[\protect\citeauthoryear{{Simpson}, {Pakmor}, {Pfrommer}, {Glover}  \&
  {Smith}}{{Simpson} et~al.}{2023}]{simpson2023cosmic}
{Simpson} C.~M.,  {Pakmor} R.,  {Pfrommer} C.,  {Glover} S. C.~O.,   {Smith}
  R.,  2023, \mn@doi [\mnras] {10.1093/mnras/stac3601}, \href
  {https://ui.adsabs.harvard.edu/abs/2023MNRAS.520.4621S} {520, 4621}

\bibitem[\protect\citeauthoryear{{Springel}}{{Springel}}{2010}]{springel2010pur}
{Springel} V.,  2010, \mn@doi [\mnras] {10.1111/j.1365-2966.2009.15715.x},
  \href {https://ui.adsabs.harvard.edu/abs/2010MNRAS.401..791S} {401, 791}

\bibitem[\protect\citeauthoryear{{Stone}, {Tomida}, {White}  \&
  {Felker}}{{Stone} et~al.}{2020}]{stone2020athena++}
{Stone} J.~M.,  {Tomida} K.,  {White} C.~J.,   {Felker} K.~G.,  2020, \mn@doi
  [\apjs] {10.3847/1538-4365/ab929b}, \href
  {https://ui.adsabs.harvard.edu/abs/2020ApJS..249....4S} {249, 4}

\bibitem[\protect\citeauthoryear{{Tsukamoto}, {Okuzumi}, {Iwasaki}, {Machida}
  \& {Inutsuka}}{{Tsukamoto} et~al.}{2017}]{tsukamoto2017impact}
{Tsukamoto} Y.,  {Okuzumi} S.,  {Iwasaki} K.,  {Machida} M.~N.,   {Inutsuka}
  S.-i.,  2017, \mn@doi [\pasj] {10.1093/pasj/psx113}, \href
  {https://ui.adsabs.harvard.edu/abs/2017PASJ...69...95T} {69, 95}

\bibitem[\protect\citeauthoryear{{Vaytet}, {Commer{\c{c}}on}, {Masson},
  {Gonz{\'a}lez}  \& {Chabrier}}{{Vaytet} et~al.}{2018}]{vaytet2018non_ideal}
{Vaytet} N.,  {Commer{\c{c}}on} B.,  {Masson} J.,  {Gonz{\'a}lez} M.,
  {Chabrier} G.,  2018, \mn@doi [\aap] {10.1051/0004-6361/201732075}, \href
  {https://ui.adsabs.harvard.edu/abs/2018A&A...615A...5V} {615, A5}

\bibitem[\protect\citeauthoryear{Velikhov}{Velikhov}{1959}]{velikhov1959stability}
Velikhov E.,  1959, Sov. Phys. JETP, 36, 995

\bibitem[\protect\citeauthoryear{{Weinberger}, {Springel}  \&
  {Pakmor}}{{Weinberger} et~al.}{2020}]{weinberger2020arepo}
{Weinberger} R.,  {Springel} V.,   {Pakmor} R.,  2020, \mn@doi [\apjs]
  {10.3847/1538-4365/ab908c}, \href
  {https://ui.adsabs.harvard.edu/abs/2020ApJS..248...32W} {248, 32}

\bibitem[\protect\citeauthoryear{Wurster}{Wurster}{2016}]{wurster2016nicil}
Wurster J.,  2016, Publications of the Astronomical Society of Australia, 33,
  e041

\bibitem[\protect\citeauthoryear{{Wurster} \& {Li}}{{Wurster} \&
  {Li}}{2018}]{wurster2018role}
{Wurster} J.,  {Li} Z.-Y.,  2018, \mn@doi [Frontiers in Astronomy and Space
  Sciences] {10.3389/fspas.2018.00039}, \href
  {https://ui.adsabs.harvard.edu/abs/2018FrASS...5...39W} {5, 39}

\bibitem[\protect\citeauthoryear{{Wurster}, {Price}  \& {Bate}}{{Wurster}
  et~al.}{2016}]{wurster2016can}
{Wurster} J.,  {Price} D.~J.,   {Bate} M.~R.,  2016, \mn@doi [\mnras]
  {10.1093/mnras/stw013}, \href
  {https://ui.adsabs.harvard.edu/abs/2016MNRAS.457.1037W} {457, 1037}

\bibitem[\protect\citeauthoryear{{Wurster}, {Bate}  \& {Price}}{{Wurster}
  et~al.}{2018}]{wurster2018outflows}
{Wurster} J.,  {Bate} M.~R.,   {Price} D.~J.,  2018, \mn@doi [\mnras]
  {10.1093/mnras/stx3339}, \href
  {https://ui.adsabs.harvard.edu/abs/2018MNRAS.475.1859W} {475, 1859}

\bibitem[\protect\citeauthoryear{{Wurster}, {Bate}  \& {Bonnell}}{{Wurster}
  et~al.}{2021}]{wurster2021}
{Wurster} J.,  {Bate} M.~R.,   {Bonnell} I.~A.,  2021, \mn@doi [\mnras]
  {10.1093/mnras/stab2296}, \href
  {https://ui.adsabs.harvard.edu/abs/2021MNRAS.507.2354W} {507, 2354}

\bibitem[\protect\citeauthoryear{{Zier} \& {Springel}}{{Zier} \&
  {Springel}}{2022a}]{zier2022simulating}
{Zier} O.,  {Springel} V.,  2022a, \mn@doi [\mnras] {10.1093/mnras/stac1783},
  \href {https://ui.adsabs.harvard.edu/abs/2022MNRAS.515..525Z} {515, 525}

\bibitem[\protect\citeauthoryear{{Zier} \& {Springel}}{{Zier} \&
  {Springel}}{2022b}]{ZierMRI}
{Zier} O.,  {Springel} V.,  2022b, \mn@doi [\mnras] {10.1093/mnras/stac2831},
  \href {https://ui.adsabs.harvard.edu/abs/2022MNRAS.517.2639Z} {517, 2639}

\makeatother
\end{thebibliography}

\section{Details of the least square fit (LSF) algorithm}
\label{app:leastSquareFit}
Our goal is to find a solution $\bm q$ that minimises the residual in equation~(\ref{eq:matrixEquationResidual}). \cite{pakmor2016semi} used for this the normal equation
\begin{equation}
    \left(\bm X \bm X^T\right) \bm q = \bm X^T \bm Y,
    \label{eq:normalEquation}
\end{equation}
which can be solved by
\begin{equation}
    \bm q = \left(\bm X \bm X^T\right)^{-1}\bm X^T \bm Y,
\end{equation}
if the matrix $\bm X \bm X^T$ is invertible. The numerical stability of this approach can be measured by the condition number $\kappa(\bm X)$ of the matrix $\bm X$, which can be defined as the ratio of the largest to the smallest singular value of $\bm X$. 

As can be seen in equation~(\ref{eq:normalEquation}), using the normal equation the condition number is given by $\kappa \left(\bm X \bm X^T\right) \approx \kappa\left(\bm X^2\right)$, which means that if the matrix ${\bm X}$ is ill-conditioned, the LSF can become unstable.
In this paper, we therefore compute a singular value decomposition (SVD) for ${\bm X}$:
\begin{equation}
    \bm X = \bm U \bm \Lambda \bm V^T ,
\end{equation}
where we have introduced the $M\times M$ orthogonal matrix $\bm U$, the $M\times N$ diagonal matrix $\bm \Lambda$, and the $N\times N$ orthogonal matrix $\bm V^T$. The diagonal components of $\bm \Lambda$ are the singular values of $\bm X$, which allows us to calculate $\kappa(\bm X)$. A closer look at (\ref{eq:componentOfX}) shows that the matrix $\bm X$ contains components without unit (first column) and with the unit of length (other columns). This means that $\kappa(\bm X)$ depends on the choice of the code unit for the length scales, which becomes particularly problematic in simulations involving different length scales. We therefore use a simple preconditioner by measuring the maximum $P = {\rm max}_{j=1,2,3} \left(\left|\bm X_{i,j}\right|\right)$ for each interface and normalising all distances in the matrix $\bm X$. This leads to condition numbers smaller than 1000 in all tests and makes it independent of the choice of the unit system. Nevertheless, for large $\kappa(\bm X) > 10000$ we reject the LSF result and use a fallback strategy to obtain the gradients (see Section~\ref{subsubsec:estimateGradients}).

The pseudoinverse is given by
\begin{equation} 
    \bm X^{-1} = \bm V \bm \Lambda^{-1} \bm U^T,
\end{equation}
where we have defined the pseudo-inverse of $\bm \Lambda$ as the diagonal matrix for which $\left(\bm\Lambda^{-1}\right)_i = 1/\Lambda_i$ holds. At the end we apply the inverse unit transformation with $P$ to get the results in code units. If we obtain a singular value of 0, the condition number becomes infinite, so we reject the LSF anyway. For computing the SVD in practice, we use the GSL library.

\section{Analytical solutions of test problems}
\label{app:analyticalSolutions}
\subsection{Diffusion of a magnetic field}
\label{app:solutionsDiffusion}
\subsubsection{Ohmic diffusion}

The solution of the differential equation (\ref{eq:diffusionEquationOhmic}) is given by 
\begin{equation}
    B_y(x, t) = \frac{B_{y0}}{\sqrt{4\pi \eta_{\rm OR} t}}\exp\left(-\frac{x^2}{4\eta_{\rm OR}t}\right)
    \label{eq:solutionDiffusionOhmic1D}
\end{equation}
for the initial conditions (\ref{eq:ICDiffusion1D}) and by
\begin{equation}
    B_y(x,z,t) = \frac{B_{y0}}{4\pi \eta_{\rm OR} t}\exp\left(-\frac{x^2+z^2}{4\eta_{\rm OR}t}\right) 
    \label{eq:solutionDiffusionOhmic2D}
\end{equation}
for the initial conditions (\ref{eq:ICDiffusion2D}).

\subsubsection{Ambipolar diffusion: Barenblatt-Pattle solution}
\label{app:barenBlattSolution}
The diffusion equation
\begin{equation}
    \frac{\partial B_y}{\partial t} = \nabla \cdot \left(B_y^\beta \nabla B_y\right)
\end{equation}
with constant $\beta$ (equal to $2$ for ambipolar diffusion) has for the initial condition of a $\delta$-peak the Barenblatt-Pattle solution:
\begin{equation}
B_y =  \left\{
\begin{array}{ll}
A t^\alpha \left[1- \left(\frac{r}{\eta_0 t^\delta} \right)^2 \right]^{1/\beta} & {\rm if }\;\; r < \eta_0 t^\delta,  \\
0 &  \textrm{if }\;\; r > \eta_0 t^\delta. \\
\end{array}
\right.
\end{equation}
Here $\mu$ is the dimensionality of the initial peak (1 or 2 in our case), and the constants are given by \citep{grundy1982eigenvalues}:
\begin{align}
    \alpha &= \frac{- \mu}{2 + \mu \beta},\\
    \delta &= \frac{1}{2 + \mu \beta},\\
    A &= \left(\frac{\delta \beta \eta_0^2}{2} \right)^{1/2},\\
    \int_r r^{\mu-1} B_{y0}\left(\bm x\right) dr &= \eta_0^{\mu + 2 / \beta} \left(\frac{\delta \beta}{2} \right)^{\beta-1}\frac{\Gamma \left(\mu/2\right)\Gamma \left(1/\beta + 1\right)}{\Gamma\left(1/\beta +1 + \mu/2\right)}.
\end{align}

\subsection{C-Shock}
\label{app:CShockEquations}

The governing equations for the oblique C-shock can be obtained from the non-ideal MHD equations by setting $v_z = B_z = 0$, and by making the assumption that all quantities are independent of the $y$- and $z$-coordinates, as well as of time. The divergence constraint then leads to $B_x = {\rm const.}$, and to 4 (5) equations for an isothermal (adiabatic) equation of state. We use the Mathematica software to solve the stationary equations with the pre-shock quantities as initial conditions, and trigger the shock by setting $\left.\frac{\partial B_y}{\partial x}\right|_{x=0} = 0.01$.

\subsubsection{Isothermal equation of state}

When we use the isothermal equation of state, $P = \rho c_s^2$, the pressure is not a free variable and the remaining equations are:
\begin{gather}
    \frac{\partial \rho v_{x}}{\partial x} = 0,\\
    \frac{\partial \left(\rho v_x^2 + \rho c_s^2 +B_y^2/2\right)}{\partial x} = 0,\\
     \frac{\partial \left(\rho v_x v_y -B_x B_y\right)}{\partial x} = 0,\\
     \frac{\partial \left(v_x B_y - v_y B_x - \frac{1}{\gamma_{\rm AD } \rho_i \ \rho}\left(B_x^2+B_y^2\right)\frac{\partial B_y}{\partial x}-\eta_{\rm OR} \frac{\partial B_y}{\partial x}\right)}{\partial x} = 0.
\end{gather}

\subsubsection{Adiabatic equation of state}
By using the adiabatic equation of state, $P =\left(\gamma -1\right) \rho u$, where we introduced the adiabatic coefficient $\gamma = 5/3$ and the thermal energy per unity mass $u$, we obtain the  system of equations:
\begin{gather}
    \frac{\partial \rho v_{x}}{\partial x} = 0,\\
    \frac{\partial \left(\rho v_x^2 +P +B_y^2/2\right)}{\partial x} = 0,\\
     \frac{\partial \left(\rho v_x v_y -B_x B_y\right)}{\partial x} = 0,\\
     \frac{\partial \left(v_x B_y - v_y B_x - \frac{1}{\gamma_{\rm AD \rho} \rho_i}\left(B_x^2+B_y^2\right)\frac{\partial B_y}{\partial x}-\eta_{\rm OR} \frac{\partial B_y}{\partial x}\right)}{\partial x} = 0,\\
      v_x \left(P \left(1 + \frac{1}{\gamma -1}\right) +B_x^2+B_y^2 + \frac{\rho}{2} \left(v_x^2+v_y^2\right)\right) \nonumber\\ -v_x Bx  - v_y B_y - \frac{1}{\gamma_{\rm AD } \rho_i \rho}\left(B_x^2+B_y^2\right)B_y\frac{\partial B_y}{\partial x}\\ - \eta_{\rm OR} B_y \frac{\partial B_y}{\partial x} = {\rm const.}\nonumber
\end{gather}

\subsection{Damped Alfvén wave}
\label{app:dampedAlfvenWave}
For the setup we use in Section~\ref{subsec:dampingAlfvenWave}, we can separate the eigenvalue into a real and imaginary part:
\begin{equation}
    s = s_r +i  s_i = -\frac{\eta_{\rm OR} k^2}{2} \pm i \sqrt{k^2 v_A^2 - \left(\frac{\eta_{\rm OR} k^2}{2} \right)^2}
\end{equation}
if we assume weak damping, which means $k^2 v_A^2  > \frac{\eta_{\rm OR}^2 k^4}{4} $. 
For this case, the amplitude of velocity perturbations is given by
\begin{equation}
    \delta v = \frac{k_z B_0}{\rho} \frac{1}{s_r^2 + s_i^2} \delta B,
\end{equation}
where $\delta B$ is the initial perturbation of the magnetic field.
The eigenvectors can now be combined to form standing or travelling waves with frequency $s_i$ and damping rate $s_r$.

\subsubsection{Travelling wave}

Travelling waves in the $z$-direction can be written as:
\begin{align}
    v_x(z,t) &= e^{s_r t} \delta v \left[s_i \cos\left(k_z z + s_i t\right) - s_r \sin\left(k_z z + s_i t\right)\right],\\
    v_y(z,t) &= -e^{s_r t} \delta v \left[s_i \sin\left(k_z z + s_i t\right) + s_r \cos\left(k_z z + s_i t\right)\right], \\
    B_x(z,t) &= e^{s_r t} \delta B \cos\left(k_z z + s_i t\right),\\
    B_y(z,t) &= -e^{s_r t} \delta B \sin\left(k_z z + s_i t\right).
\end{align}

\subsubsection{Standing wave}

The time evolution of standing waves is given by:
\begin{align}
    v_x(z,t) &= -e^{s_r t} \delta v \sin\left(k_z z\right) \left[s_i  \sin\left(s_i t\right)+ s_r \cos\left(s_i t\right)\right],\\
    v_y(z,t) &= -e^{s_r t} \delta v \cos\left(k_z z\right) \left[s_i  \sin\left(s_i t\right)+ s_r \cos\left(s_i t\right)\right], \\
    B_x(z,t) &= e^{s_r t} \cos\left(s_i t\right) \delta B \cos\left(k_z z \right)),\\
    B_y(z,t) &= -e^{s_r t} \cos\left(s_i t\right) \delta B \sin\left(k_z z\right).
\end{align}

\begin{figure*}
    \centering
    \includegraphics[width=1\linewidth]{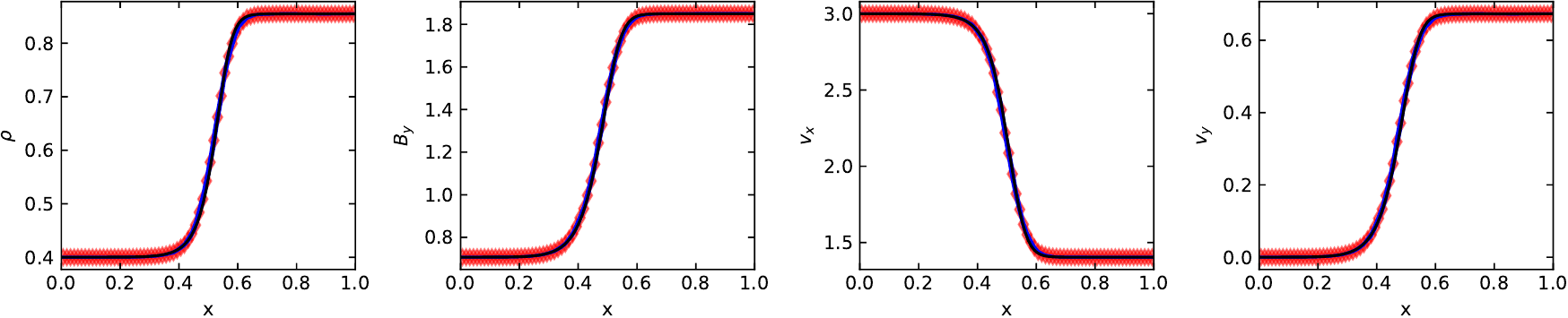}
    \caption{The structure of an isothermal, oblique C-shock with Ohmic diffusion and sound speed $c_s = 1$. The black line corresponds to the semi-analytically expected structure of the shock while the red symbols show the average values in a simulation with {\small AREPO} at $t=5$ and on average 40 cells per dimension (we use 80 bins in the $x$-direction for this plot). The blue lines show the $\pm 1$ standard deviation in each bin. More details can be found in Section~\ref{subsubsec:CShockOhmic}.}
    \label{fig:OhmicIsothermal_t5_40_cells}
\end{figure*}

\begin{figure*}
    \centering
    \includegraphics[width=1\linewidth]{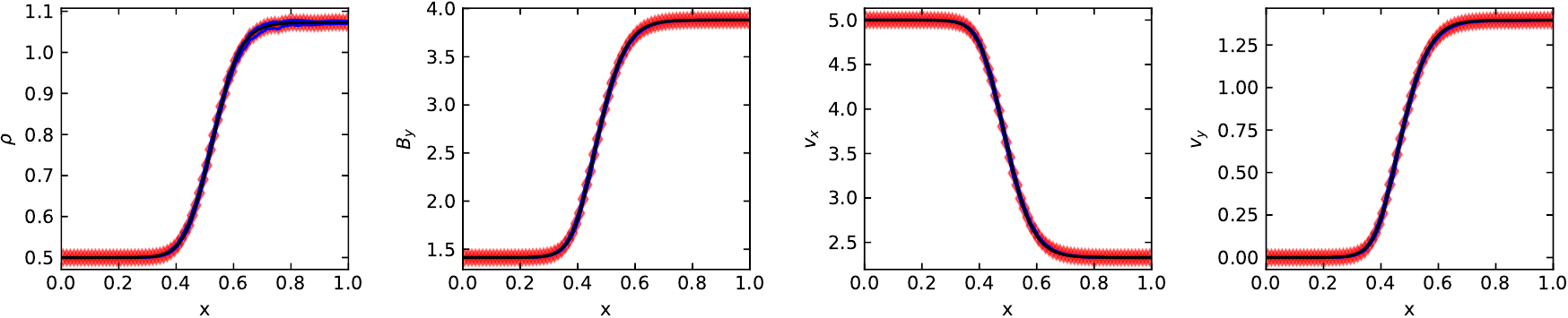}
    \caption{The structure of an isothermal, oblique C-shock with ambipolar diffusion and sound speed $c_s = 0.5$. The black line corresponds to the semi-analytic expected structure of the shock while the red symbols show the average values in a simulation with {\small AREPO} at $t=5$ and on average 40 cells per dimension (we use 80 bins in the $x$-direction for this plot). The blue lines show the values $\pm 1$ standard deviation in each bin. More details can be found in Section~\ref{subsubsec:CShockAmbipolar}.}
    \label{fig:AmbipolarIsothermal_t5_40_cells}
\end{figure*}

\section{Results for the isothermal C-shock test}
\label{app:resultsIsothermalCShock}

We have repeated the C-shock tests presented in Section~\ref{subsec:CShock} with an isothermal equation of state $p= \rho c_s^2$ and an isothermal sound speed $c_s = 1$ for the Ohmic diffusion case and $c_s = 0.5$ for ambipolar diffusion. In \cref{fig:OhmicIsothermal_t5_40_cells} and \cref{fig:AmbipolarIsothermal_t5_40_cells} we show the structure of the shock in our simulations, and in \cref{fig:L1_error_CShock_Ohmic_Isothermal} and \cref{fig:L1_error_CShock_Ambipolar_Isothermal} the $L_1$ error norm as a function of resolution.  As for the non-isothermal case, {\small AREPO} is able to accurately capture the structure of the shock and shows good convergence behaviour. The noise in the density is much smaller since due to the isothermal equation of state this noise would automatically lead to noise in the pressure distribution. However, pressure differences would affect the dynamics and will be washed out. Nevertheless, we can still see a small amount of noise which translates into slightly larger errors in the density compared to {\rm ATHENA++}.

\setcounter{figure}{2}   
\begin{figure}
    \centering
    \includegraphics[width=1\linewidth]{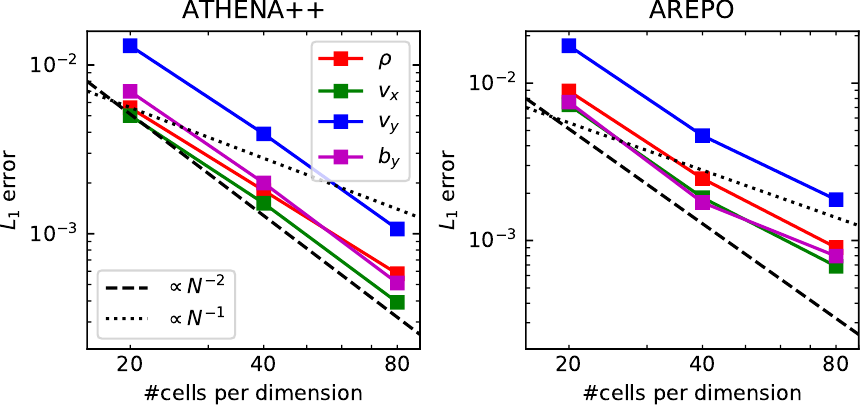}
    \caption{The $L_1$ error norm as a function of resolution for the isothermal C-shock test with Ohmic diffusion. We show results obtained with {\small ATHENA++} (left) and {\small AREPO} (right). More details can be found in Section~\ref{subsubsec:CShockOhmic}.}
    \label{fig:L1_error_CShock_Ohmic_Isothermal}
\end{figure}
    
\begin{figure}
    \centering
    \includegraphics[width=1\linewidth]{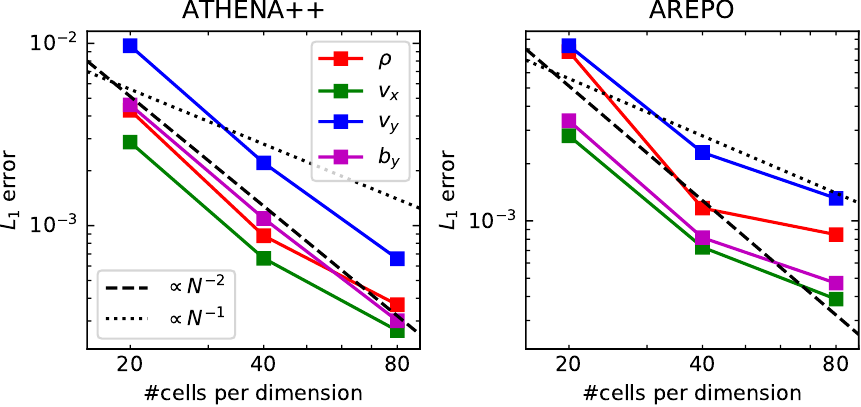}
    \caption{The $L_1$ error as a function of resolution for the isothermal C-shock test with ambipolar diffusion. We show results obtained with {\small ATHENA++} (left) and {\small AREPO} (right).
    More details can be found in Section~\ref{subsubsec:CShockAmbipolar}.}
    \label{fig:L1_error_CShock_Ambipolar_Isothermal}
\end{figure}

 \section{The linearized equations for the isothermal resistive MRI}
 \label{app:fullDispersionRelatio}
 The starting point of our analysis is equation~(\ref{eq:shearingBoxEquations}), which we linearize around the ground state and add the perturbations described in Section~\ref{subsec:nonidealMHDDispersion}. We can use the condition $\nabla \cdot \bm b = 0$ to eliminate $b_z$ by $b_z = -k_z / k_x b_x$, and the equation of state to eliminate the pressure perturbations. We will use $\Omega_0 = 1$. Then the linearized equations can be written as an eigenvalue problem:
 \begin{equation}
 \sigma \bm V = A_{\rm id} \bm V + A_{\rm OR} \bm V + A_{\rm AD} \bm V,
 \label{eq:eigenvalueProblem}
 \end{equation}
where the eigenvector $V = \left(\rho,v_x,v_y, v_z, b_x, b_y\right)^T$ contains the amplitude of the perturbations and the eigenvalue $\sigma$ is equivalent to the growth rate. The matrices containing the linearized terms depend on the components $k_x$ and $k_z$ of the wave vector of the perturbation, the background density $\rho_0$, the components $B_y$ and $B_z$ of the background magnetic field, the shear parameter $q$, and the strength of the non-ideal MHD terms. The matrix containing the ideal MHD terms can be written as:
 \begin{equation}
 \resizebox{0.99\hsize}{!}{
     $A_{\rm id} = \begin{pmatrix}
       0& -i\rho k_x &0& -i\rho_0 k_z&0&0\\
       \frac{-i c_s^2 k_x}{\rho_0} &0& 2&0&\frac{i k_z B_z + ik_x^2 B_Z / k_z}{ \rho_0}& \frac{-i B_y k_x}{\rho_0}\\
       0 &q-2 & 0 & 0&0& \frac{i k_z B_z}{ \rho_0}\\
        \frac{-i c_s^2k_z}{\rho_0} &0 &0 &0 &0& \frac{-i B_y k_z}{\rho_0}\\
       0 &ik_z B_z & 0 &0 &0&0\\
       0 & -i k_x B_y & i k_z B_z & -i k_z B_y &-q&0
      \end{pmatrix}$}.
 \end{equation}
 The non-ideal terms only influence the evolution of the magnetic field and can be written as:
 \begin{equation}
     A_{\rm OR} = \begin{pmatrix}
     0 &\cdots &0&0\\
     \vdots & \ddots  &\vdots&\vdots\\
     0 &0  & -\frac{k^2 v_A^2}{\Lambda_{\rm OR}} & 0\\
     0 &0 & 0 & -\frac{k^2 v_A^2}{\Lambda_{\rm OR}}\\
     \end{pmatrix}
 \end{equation}
and
   \begin{equation}
     A_{\rm AD} = \begin{pmatrix}
     0 &\cdots &0&0\\
     \vdots & \ddots  &\vdots&\vdots\\
     0 &0  & -\frac{k^2 v_A^2}{\Lambda_{\rm AD}} \frac{B_z^2}{B^2} &  \frac{k_x k_z v_A^2}{\Lambda_{\rm AD}} \frac{B_y B_z}{B^2}\\
     0 &0 & \frac{v_A^2}{\Lambda_{\rm AD}}\frac{B_y B_z}{B^2} \frac{k_x k^2}{k_z}& - \frac{v_A^2}{\Lambda_{\rm AD}} \left(k_z^2\frac{B_z^2}{B^2} + k^2 \frac{B_y^2}{B^2}\right) \\
     \end{pmatrix}.
 \end{equation}
Equation~(\ref{eq:eigenvalueProblem}) can be solved using standard linear algebra software. We then choose the eigenvalue with the largest real value, which corresponds to the fastest-growing mode. We note that our equations are slightly different to those often analyzed in the literature \citep[e.g. ][]{kunz2004ambipolar, lesur2021magnetohydrodynamics}, which assume incompressibility instead of an isothermal equation of state. In this case, the continuity equation simplifies to $\nabla \cdot \bm v = 0$, which allows additionally the removal of $v_z$ from the linear system. The pressure in this case is set so that $\nabla \cdot \bm v = 0$ holds. This reduces the equation system to only containing the independent variables $v_x$, $v_y$, $b_x$ and $b_y$ which can be more easily analyzed analytically. We compared the solutions of our system with those from the incompressible model and found that the eigenvalues are very similar (even the same for $k_x = 0$), but for oblique modes, the eigenvectors change and contain density perturbations. This is crucial to obtain the best initial conditions for our code tests.

\subsection{Numerical data of the eigenvectors used in simulations of the linear MRI}

To improve the reproducibility of our tests we present in this section the exact values of the eigenvectors we use.
We found that they can be written in general as:
\begin{align} 
    b_x = b_0 \cos\left(\phi_B\right)  \cos\left(k_x x + k_z z\right),\\
    b_y = b_0 \sin\left(\phi_B\right)  \cos\left(k_x x + k_z z\right),\\
     v_x = -\frac{v_0}{b_0} b_0 \cos\left(\phi_v\right) \sin\left(k_x x + k_z z\right),\\
     v_y = -\frac{v_0}{b_0} b_0 \sin\left(\phi_v\right) \sin\left(k_x x + k_z z\right),\\
     v_z = -\frac{v_0}{b_0} b_0  \frac{v_{z0}}{v_{0}} \sin\left(k_x x + k_z z\right),\\
      \rho = \frac{\rho_0}{v_0} \frac{v_0}{b_0} b_0  \cos\left(k_x x + k_z z\right),
\end{align}
where $b_0$ is the arbitrary initial amplitude of the perturbation. In \cref{tab:linearGrwothOhmic} we give the results of the remaining constants for Ohmic diffusion with $k_x = 0$ and $k_z = 2\pi$, with a background field defined by the plasma $\beta = 84.2199$. In this case we find $\rho_0 = 0$ from which $v_z = 0$ follows.

For the case of ambipolar diffusion, we use $k_x = -2\pi = -k_z$ and a background field with $B_y = B_z$. The background magnetic field is defined by the plasma $\beta = 169.84 = B_z^2 / \rho_0$.
In this case, there are also density perturbations in the eigenvector and we show in \cref{tab:linearGrowthAmbipolar} the full eigenvector.
\begin{table}

    \centering
    \begin{tabular}{c|c|c|c|c|c}
    \hline
        $\Lambda_0$ & $\sigma$ & $\phi_B$ & $\phi_v$ & $v_0/b_0$\\
        \hline
        \hline
0.8 & 0.11108 & -0.64237 & -2.74319 & 1.15109  \\
1.0 & 0.22910 & -0.65580 & -2.66494 & 1.07473  \\
1.5 & 0.39076 & -0.68672 & -2.56041 & 0.97065  \\
2.0 & 0.47524 & -0.70726 & -2.50829 & 0.91939  \\
3.0 & 0.56298 & -0.73082 & -2.45662 & 0.86939  \\
5.0 & 0.63585 & -0.75158 & -2.41586 & 0.83055  \\
10.0 & 0.69217 & -0.76813 & -2.38577 & 0.80221  \\
$\infty$ & 0.75000 & -0.78539 & -2.35618 & 0.77460\\  
        \hline
    \end{tabular}
    \caption{The properties of the eigenvectors of the linear MRI with $k_x = 0$, $\Omega_0 =1$, $c_s =1$, $\rho=1$, and only Ohmic diffusion. 
    See Appendix~\ref{app:fullDispersionRelatio} for details.}
    \label{tab:linearGrwothOhmic}
\end{table}
 
\begin{table}
\resizebox{\linewidth}{!}{%
    \centering
    \begin{tabular}{c|c|c|c|c|c|c|c}
    \hline
        $\Lambda_{\rm AD}$ & $\sigma$ & $\phi_B$ & $\phi_v$ & $v_0/b_0$ & $v_{z0}/v_0$ & $\rho_0 / v_0$\\
        \hline
        \hline
0.6 & 0.09916 & 2.34086 & 0.57363 & 1.03311 & 0.83854 & -0.08864 \\
0.8 & 0.19564 & 2.32863 & 0.68697 & 0.96980 & 0.76989 & -0.10525 \\
1.0 & 0.25520 & 2.31627 & 0.75269 & 0.92882 & 0.72517 & -0.11530 \\
1.5 & 0.33873 & 2.29139 & 0.83505 & 0.87329 & 0.66418 & -0.12918 \\
3.0 & 0.42915 & 2.25193 & 0.90540 & 0.81959 & 0.60751 & -0.14435 \\
5.0 & 0.46777 & 2.22999 & 0.92769 & 0.79948 & 0.58844 & -0.15110 \\
10.0 & 0.49778 & 2.21030 & 0.94117 & 0.78522 & 0.57644 & -0.15656 \\
$\infty$ & 0.52874 & 2.18714 & 0.95115 & 0.77187 & 0.56708 & -0.16244 \\
        \hline
    \end{tabular}}
    \caption{The properties of the eigenvectors of the linear MRI with $k_x = -k_z$, $\Omega_0 =1$, $c_s =1$, $\rho=1$, and only ambipolar diffusion.
     See Appendix~\ref{app:fullDispersionRelatio} for details.}
    \label{tab:linearGrowthAmbipolar}
\end{table}

\end{appendix}
\bsp	
\label{lastpage}
\end{document}